\documentclass[%
reprint,
superscriptaddress,
showpacs,
preprintnumbers,
nofootinbib,
amsmath,
amssymb,
amsfonts,
aps,
prd,
floatfix,
showkeys
]{revtex4-2}

\RequirePackage{atbegshi}

\usepackage[utf8]{inputenc}
\usepackage[T1]{fontenc}
\usepackage[ngerman,english]{babel}
\usepackage[matha]{mathabx} 
\usepackage[dvipsnames, svgnames, table]{xcolor}
\usepackage[braket, qm]{qcircuit}
\usepackage{%
    graphicx,
    siunitx,
    multirow,
    booktabs,
    bbold,    
    pifont,   
    capt-of,
    pgfplotstable,
    todonotes,
}
\usepackage{subfigure}     
\renewcommand{\thesubfigure}{(\alph{subfigure})}
\makeatletter
  \renewcommand{\@thesubfigure}{\thesubfigure\space}
  \def\@currentlabel{\p@subfigure\thesubfigure}
\makeatother
\usepackage{hyperref}
\usepackage{braket}
\hypersetup{
    colorlinks=true,
    citecolor=black,
    linkcolor=black,
    urlcolor=black,
    anchorcolor=black,
    linktocpage
}
\usepackage{outlines}
\usepackage{etoolbox} 
\usepackage{cleveref} 
\makeatletter
\appto{\appendix}{%
  \@ifstar{\def\theequation@prefix{A.}}%
          {}%
}
\makeatother
\crefname{figure}{Figure}{Figures}
\crefname{table}{Table}{Tables}
\crefname{equation}{Eq.}{Eqs.}
\crefname{section}{Section}{Sections}
\crefformat{appendix}{the #2Appendix#1#3} 
\graphicspath{{./figures/}}

\DeclareMathOperator{\Tr}{Tr}

\newcommand{\clqcd}{CL\kern-.25em\textsuperscript{2}QCD}
\newcommand{\vev}[1]{ \langle \, #1 \, \rangle }
\newcommand{\bpsi}{\bar{\psi}}

\newcommand{\beq} {\begin{eqnarray}}
\newcommand{\eeq} {\end{eqnarray}}
\newcommand{\nn}{ \nonumber}

\begin{document}

\title{
Simulating $\mathbb{Z}_2$ Lattice Gauge Theory with the Variational Quantum Thermalizer
}

\author{Michael Fromm}
 \email{mfromm@itp.uni-frankfurt.de}
 \affiliation{
  Institut f\"{u}r Theoretische Physik, Goethe-Universit\"{a}t Frankfurt\\
 Max-von-Laue-Str.\ 1, 60438 Frankfurt am Main, Germany
}

\author{Owe Philipsen}
 \email{philipsen@itp.uni-frankfurt.de}
 \affiliation{
  Institut f\"{u}r Theoretische Physik, Goethe-Universit\"{a}t Frankfurt\\
 Max-von-Laue-Str.\ 1, 60438 Frankfurt am Main, Germany
}

\author{Michael Spannowsky}
 \email{michael.spannowsky@durham.ac.uk}
 \affiliation{
University of Durham, Science Laboratories\\
South Rd, Durham DH1 3LE, UK
}

\author{Christopher Winterowd}
\email{winterowd@itp.uni-frankfurt.de}
\affiliation{
 Institut f\"{u}r Theoretische Physik, Goethe-Universit\"{a}t Frankfurt\\
 Max-von-Laue-Str.\ 1, 60438 Frankfurt am Main, Germany
}

\begin{abstract}
The properties of strongly-coupled lattice gauge theories at finite density as well as in real time have largely eluded first-principles studies on the lattice. This is due to the failure of importance sampling for systems with a complex action. An alternative to evade the sign problem is quantum simulation. Although still in its infancy, a lot of progress has been made in devising algorithms to address these problems. In particular, recent efforts have addressed the question of how to produce thermal Gibbs states on a quantum computer. In this study, we apply a variational quantum algorithm to a low-dimensional model which has a local abelian gauge symmetry. We demonstrate how this approach can be applied to obtain information regarding the phase diagram as well as unequal-time correlation functions at non-zero temperature.
\end{abstract}

\preprint{IPPP/23/25}

\maketitle

\section{Introduction}

Much interest has been generated recently regarding the potential of quantum computing and quantum simulation to address long-standing problems in high-energy physics \cite{Banuls:2019bmf,Bauer:2022hpo}. Traditionally, lattice gauge theory has been applied in the Euclidean formulation, where advances in computing power have led to great success in studying properties of strongly-interacting matter in thermal equilibrium. However, due to the notorious sign problem, there has been a growing interest in the Hamiltonian formulation and its application to address non-zero chemical potential and real-time dynamics. For digital quantum computers in the medium term, one problem that has arisen is how to produce thermal states.

In recent years, there has already been an explosion of progress in tackling this problem. Each approach to producing a thermal state has had to address the issue of how to produce a mixed state. This can typically be done by enlarging the system size and then entangling the ancillary system with the original system. One idea which takes this approach to create a Gibbs state via purification is the so-called thermofield double states method~\cite{Wu2018}. Another interesting method is known as active cooling \cite{Ball2022}. This takes a general initial state, couples it to a thermal heat bath, and uses the concept of the ``Maxwell demon" to obtain a thermal state at the desired temperature. Alternatively, as the Boltzmann operator is non-unitary, one can also attempt to directly approximate its action on quantum states. Using judiciously chosen initial states produced by Haar-random circuits, one can use this approximation of the Boltzmann operator to obtain Gibbs states for spin models \cite{Powers2021} as well as gauge theories \cite{Davoudi2022}.

\begin{figure*}[t]
\includegraphics[scale=.5]{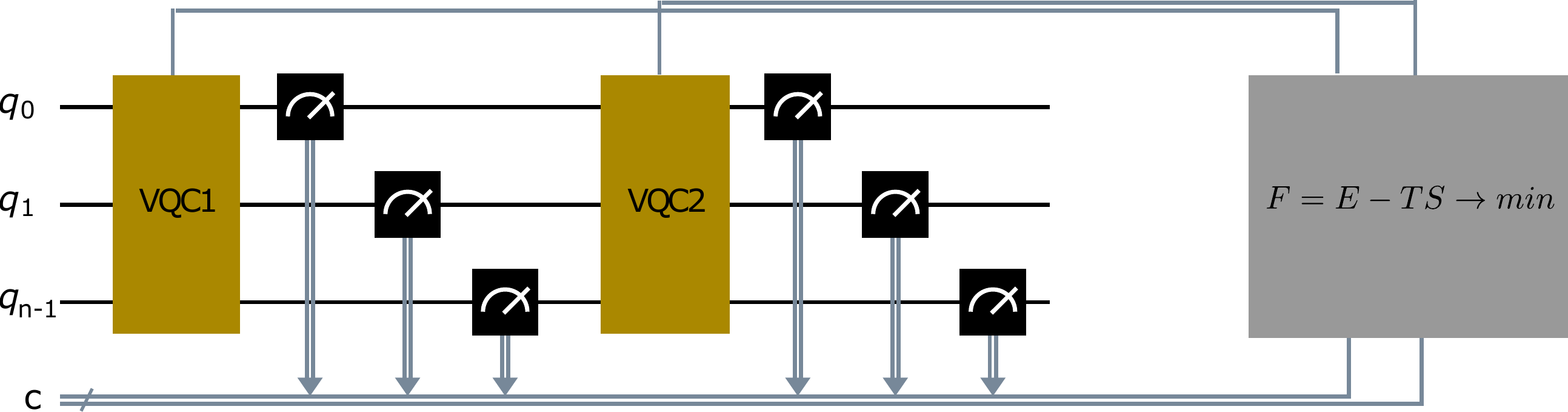}
\caption{\label{fig:vqt_sketch} Schematic depiction of the hybrid approach used in the VQT algorithm. The action of the variational circuits $VQC_1$ and $VQC_2$ and the measurements are performed on the quantum device. The results of the measurements are then sent to the classical device, which is depicted in grey. This information allows one to construct the cost function and propose a new value for the variational parameters using one's optimizer of choice.}
\end{figure*}

One can also study the problem of producing thermal states using variational methods. One such method, the variational quantum thermalizer (VQT), introduced in~\cite{Verdon2019}, serves as a natural extension of the variational quantum eigensolver (VQE)~\cite{Peruzzo2014, McClean2015} to the case of non-zero temperature. The latter uses the variational principle at $T = 0$ to approximate the ground-state of a Hamiltonian $\hat{H}$ by a parametrized ansatz $\ket{\psi_{\boldsymbol\xi}}$, subject to
\beq\label{eq:zeroTvarpriniciple}
E_0 \leq \bra{\psi_{\mathbf{\boldsymbol\xi}}} \hat{H} \ket{\psi_{\mathbf{\boldsymbol\xi}}}. 
\eeq
Similarily, for the case of $T\neq 0$, VQT addresses the problem of finding a variational approximation $\hat\rho_{var}$ to the thermal state, $\hat\rho_{Gibbs} \sim e^{-\beta \hat{H}}$ for inverse temperature $\beta = 1/T$. Similar to its zero temperature analogue in Eq.(\ref{eq:zeroTvarpriniciple}), this problem can also be reformulated as an optimization problem for the free energy with respect to variational parameters $\boldsymbol\xi$, such that
\beq\label{eq:nonzeroTvarpriniciple}
F \equiv \min_{\mathbf{\boldsymbol\xi}} F(\mathbf{\boldsymbol\xi}) = \min_{\mathbf{\boldsymbol\xi}} \left( \mathrm{tr}(\hat H\hat\rho_{\boldsymbol\xi}) + T\mathrm{tr}(\hat\rho_{\boldsymbol\xi} \log{\hat\rho_{\boldsymbol\xi}}) \right)\, ,
\eeq
 where the second term on the right-hand side involves the von Neumann entropy $S = -\mathrm{tr}(\hat\rho\log{\hat\rho})$. As the density operator appears non-linearly in the expression for the entropy and thus is not simply determined by a quantum-mechanical expectation value, several avenues have been taken in the literature: The original formulation of VQT~\cite{Verdon2019} employs classical sampling of the distribution corresponding to $\rho_{\boldsymbol\xi}$ which necessitates a model ansatz. Through this ansatz, the entropy is then given \emph {classically} by a closed-form expression. To prepare the latent distribution corresponding to $\rho_{var}$ in a \emph{quantum} circuit, the noise-assisted variational quantum thermalizer~\cite{Foldager2021}  (NAVQT) uses (simulated) parameterized depolarization gates that allow for the preparation of a mixed state, again with a closed-form, approximate expression for the entropy in terms of the noise level $\lambda$, which thus becomes a variational parameter. Of particular interest to us, due to both its simplicity and versatility, is a variant of the VQT introduced in~\cite{Selisko2022}, see Fig.\ref{fig:vqt_sketch}. While the first two examples used a classical sampling of input states~\cite{Verdon2019} or stochastic mixtures of unitary circuits~\cite{Foldager2021}, respectively, to create mixed quantum states, the approach in \cite{Selisko2022} utilizes intermediate projective measurements. As measurements are non-unitary operations, they act as a source of entropy, ultimately leading to a variational approximation of the thermal state\footnote{For a version of this algorithm using entanglement of the system with $n_q$ ancillary qubits followed by a projective measurement on the latter, see~\cite{Consiglio2023}. }.

 The subject of our study is the Hamiltonian formulation of the $\mathbb{Z}_2$ lattice gauge theory (LGT) in $1+1d$ ($\mathbb{Z}_2^{1+1}$), in which we investigate the application of the VQT to this theory. This is performed in its formulation with a gauge-redundant Hilbert space~\cite{Horn1979} and its recent, resource-efficient form derived systematically in~\cite{ZoharZ22022, Irmejs2022}. In addition, we investigate several observables of interest at both non-zero temperature $T$ and non-zero chemical potential $\mu$. 

\begin{figure*}[th]
\centering
\includegraphics[scale=0.7]{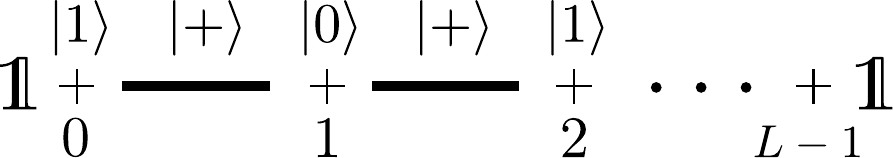}
\caption{\label{fig:initial_state} Depiction of a gauge-invariant state for $Z_2^{1+1}$ LGT with staggered fermions on a lattice with linear extent $L$. The links are all in the eigenstate of $X$ with eigenvalue $+1$, the even sites are empty (fermions), and the odd sites are filled (anti-fermions). We note here that the links at the boundary are set to unity.}
\end{figure*}
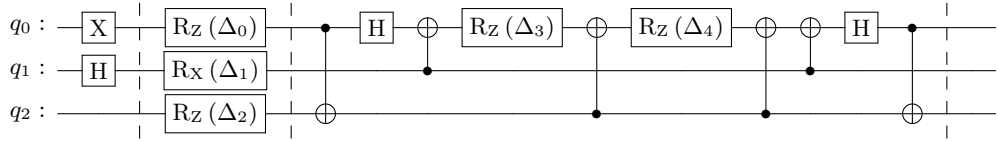
\begin{figure*}[th]
\scalebox{1.0}{
\Qcircuit @C=1.0em @R=0.2em @!R { \\
	 	\nghost{{q}_{0} :  } & \lstick{{q}_{0} :  } & \gate{\mathrm{X}} \barrier[0em]{2} & \qw & \gate{\mathrm{R_Z}\,(\mathrm{{\ensuremath{\Delta_0}}})} \barrier[0em]{2} & \qw & \ctrl{2} & \gate{\mathrm{H}} & \targ & \gate{\mathrm{R_Z}\,(\mathrm{{\ensuremath{\Delta_3}}})} & \targ & \gate{\mathrm{R_Z}\,(\mathrm{{\ensuremath{\Delta_4}}})} & \targ & \targ & \gate{\mathrm{H}} & \ctrl{2} \barrier[0em]{2} & \qw & \qw & \qw\\
	 	\nghost{{q}_{1} :  } & \lstick{{q}_{1} :  } & \gate{\mathrm{H}} & \qw & \gate{\mathrm{R_X}\,(\mathrm{{\ensuremath{\Delta_1}}})} & \qw & \qw & \qw & \ctrl{-1} & \qw & \qw & \qw & \qw & \ctrl{-1} & \qw & \qw & \qw & \qw & \qw\\
	 	\nghost{{q}_{2} :  } & \lstick{{q}_{2} :  } & \qw & \qw & \gate{\mathrm{R_Z}\,(\mathrm{{\ensuremath{\Delta_2}}})} & \qw & \targ & \qw & \qw & \qw & \ctrl{-2} & \qw & \ctrl{-2} & \qw & \qw & \targ & \qw & \qw & \qw\\
}}
\caption{\label{fig:Horn_varcirc} Depiction of a single-layer of a variational quantum circuit in the gauge-redundant formulation. The layer is composed of one sublayer of single-qubit rotation gates and a sublayer containing the decomposed three-qubit gate which is formed from the gauge-matter term in the gauge-redundant formulation. All relevant gates contain variational parameters $\Delta_i$.}
\end{figure*}
\section{Background}
\subsection{Variational Quantum Thermalizer}
As displayed in Fig.\ref{fig:vqt_sketch}, starting from an $n_q$-qubit initial state $|\psi_0\rangle$ with density matrix $\rho_0 = |\psi_0\rangle \langle \psi_0 |$, we set up a concatenation of two separate variational circuits $\mathrm{VQC}_1$ and $\mathrm{VQC}_2$. The first variational circuit, $\mathrm{VQC}_1$, creates the state $|\psi_{\mathrm{VQC}_1}\rangle$, which is given by 
\beq
\ket{\psi_{\mathrm{VQC}_1}} \equiv U_{1}(\boldsymbol\theta) \ket{\psi_0} = \sum_{i=0}^{2^{n_q} - 1} a_i(\boldsymbol\theta)\ket{i}\,.
\eeq
An intermediate measurement in the computational basis collapses this state and statistically yields the latent density matrix
\beq
 \rho_{1}(\boldsymbol\theta) =  \sum_{i=0}^{2^{n_q} - 1} |a_i(\boldsymbol\theta)|^2 \ket{i} \bra{i}\,.
\eeq
The subsequent application of the variational circuit $\mathrm{VQC}_2$ transforms this into
\beq \nn
\label{eq:rho_var}
 \rho_{2}(\boldsymbol\theta, \boldsymbol\phi) &=&  U_{2}(\boldsymbol\phi)  \rho_{1}(\boldsymbol\theta)U^\dagger_{2}(\boldsymbol\phi), \\ \label{eq:selisko_rho} &\equiv& \sum_{i=0}^{2^{n_q} - 1} p_i(\boldsymbol\theta) \ket{\psi_i(\boldsymbol\phi)} \bra{\psi_i(\boldsymbol\phi)},
\eeq
 after which a measurement of the energy is performed. The von Neumann entropy is then calculated as \\ $S = -\sum_{i} p_i \log{p_i}$, with  $p_i \approx n_i/n_s$, where $n_i$ represents the counts for state $i$ in the computational basis and $n_s$ represents the number of shots. By varying the parameter set $\boldsymbol\xi = (\boldsymbol\theta, \boldsymbol\phi)$ in an optimal way, one obtains a variational approximation to the Gibbs state $\hat\rho_{Gibbs} \sim e^{-\beta \hat H}$. The fact that the circuit is broken up into two parts is intended for noisy intermediate-scale quantum (NISQ) devices, as the depth of each variational circuit can be kept relatively shallow. This approach, however, naturally leads to a truncation in the number of states in Eq.(\ref{eq:selisko_rho}), as both the number of shots, $n_s$, and the memory necessary to store the histogram describing the counts scales with the dimension of the Hilbert space of the underlying problem. We pursue this discussion in further detail in Sect.\ref{subsect:resources}. In~\cite{Selisko2022}, the algorithm was demonstrated to reach convergence for the transverse field Heisenberg-Model in $d = 1,2$. Applying qVQT to gauge theories will introduce the need for a gauge-invariant algorithm implementation. We show below for the case of $\mathbb{Z}_2^{1+1}$ how this can be achieved.
 
\subsection{$\mathbb{Z_2}$ LGT in $1+1d$}
This section briefly outlines our two approaches to $\mathbb{Z}_2$ lattice gauge theory in the Hamiltonian formulation. In the first approach, introduced in \cite{Horn1979} and referred to as the gauge-redundant formulation, local gauge invariance must be enforced on the states to define the physical Hilbert space. This can be achieved, for example, by introducing a penalty term as in \cite{Davoudi2022}. In the second approach, used in \cite{ZoharZ22022} and referred to as the resource-efficient formulation, the matter states are decoupled, and the constraint imposed by Gauss' law is built into the Hamiltonian.
\subsubsection{Gauge-Redundant Formulation}
In the so-called group basis, $\mathbb{Z}_2^{1+1}$ LGT with a single flavour of staggered fermion is formally given by the Hamiltonian $ H =  H_E + H_{GM} + H_M + H_{\mu}$ with
\beq
\label{eq:Z2horn}
H_E &=& -\epsilon \sum_{n=0}^{L-2} X_n,\\
H_{GM} &=& \frac{J}{2}\sum_{n}^{L-2} \left(\sigma^x_n Z_n \sigma^x_{n+1} + \sigma^y_n Z_n \sigma^y_{n+1} \right),\\
H_M &=& \frac{m}{2} \sum_{n=0}^{L-1} (-1)^n \sigma_n^z,\\
H_\mu &=& -\frac{\mu}{2}\sum_{n=0}^{L-1} \sigma_n^z\,.
\eeq
In the above equations, the fermionic creation and annihilation operators have been replaced by Pauli matrices living at the sites of the lattice via the usual Jordan-Wigner transformation.
Here we have adopted the familiar convention of denoting Pauli operators acting on the link Hilbert space of the gauge bosons by $\{X,Y,Z\}$, and those acting on the fermionic matter space by $\{\sigma^x, \sigma^y, \sigma^z\}$. The occurrence of both gauge-bosonic and fermionic degrees of freedom implies a gauge redundant Hilbert space. With our choice of basis, a gauge transformation at site $n$ is defined by~\cite{Zohar2014}
\beq\label{eq:z2_gauge_trafo}
{\Theta}_g(n) &=& X_nX_{n-1} e^{i\pi \left[\psi^{\dagger}_n\psi_n + \frac{1}{2}((-1)^n-1) \right]}, \nonumber\\
&=& (-1)^{n+1}X_nX_{n-1} \sigma_n^z\,.
\eeq
We illustrate a gauge-invariant state $|\psi\rangle$ of the lattice system in Fig.\ref{fig:initial_state}, where the fermionic states are in the computational basis, and the gauge bosons are in the $X$-eigenbasis, $X\ket{\pm} = \pm \ket{\pm}$. This subspace of the total Hilbert space is needed to compute physical observables.
\subsubsection{Resource-Efficient Formulation}
It turns out that one can eliminate the fermionic matter fields while simultaneously removing the Hilbert space redundancy of the theory. This is accomplished in a two-step process whereby the fermions are first mapped to hard-core bosons by a unitary transformation and are then decoupled from the gauge bosons by a second unitary transformation. This method is, in fact, a very powerful formalism which can also be applied in arbitrary dimensions to continuous gauge groups such as $SU(N)$ and $U(N)$ \cite{Zohar2018}. For $\mathbb{Z}_2$ LGT in $1+1$ dimensions, the Hamiltonian in this formulation is as follows
\beq
\label{eq:H_Zohar_Z2_static_background_fixed}
H^{(2)}_E &=& - \sum_{n=0}^{L-2}\left( \epsilon Z_n + \frac{J}{2}Y_n\right),\\ \nn
H^{(2)}_{GM} &=& -\frac{J}{2}\sum_{n = 1}^{L-3} Z_{n-1} Y_n Z_{n+1} \\ && -\frac{J}{2}\left(Y_0 Z_1 + Z_{L-3}Y_{L-2}\right),\label{eq:H2GM}\\
H^{(2)}_M &=& -\frac{m}{2} \sum_{n=1}^{L-1} Z_{n-1}Z_n -\frac{m}{2}\left(Z_0 + Z_{L-2}\right), \label{eq:H2M}\\ \nn 
H^{(2)}_\mu &=& \frac{\mu}{2}\sum_{n=1}^{L-2} \varepsilon_n Z_{n-1}Z_n  \\ &&+\frac{\mu}{2}\left(Z_0 +\varepsilon_{L-1} Z_{L-2}\right)\,.
\eeq
One notices that the Hamiltonian obtained by performing the above-mentioned transformations only vaguely resembles the original $1+1$-dimensional Kogut-Susskind Hamiltonian. For example, the gauge-matter interaction in eq. (\ref{eq:H2GM}) couples sets of three adjacent links. Similarly, for the term originating from the staggered mass in Eq. (\ref{eq:H2M}), one obtains a coupling between two adjacent links. With this resource-efficient formulation, one can simulate larger systems without worrying about the elements of the variational circuit preserving gauge invariance.  
\begin{figure*}[t]
\includegraphics[scale=.5]{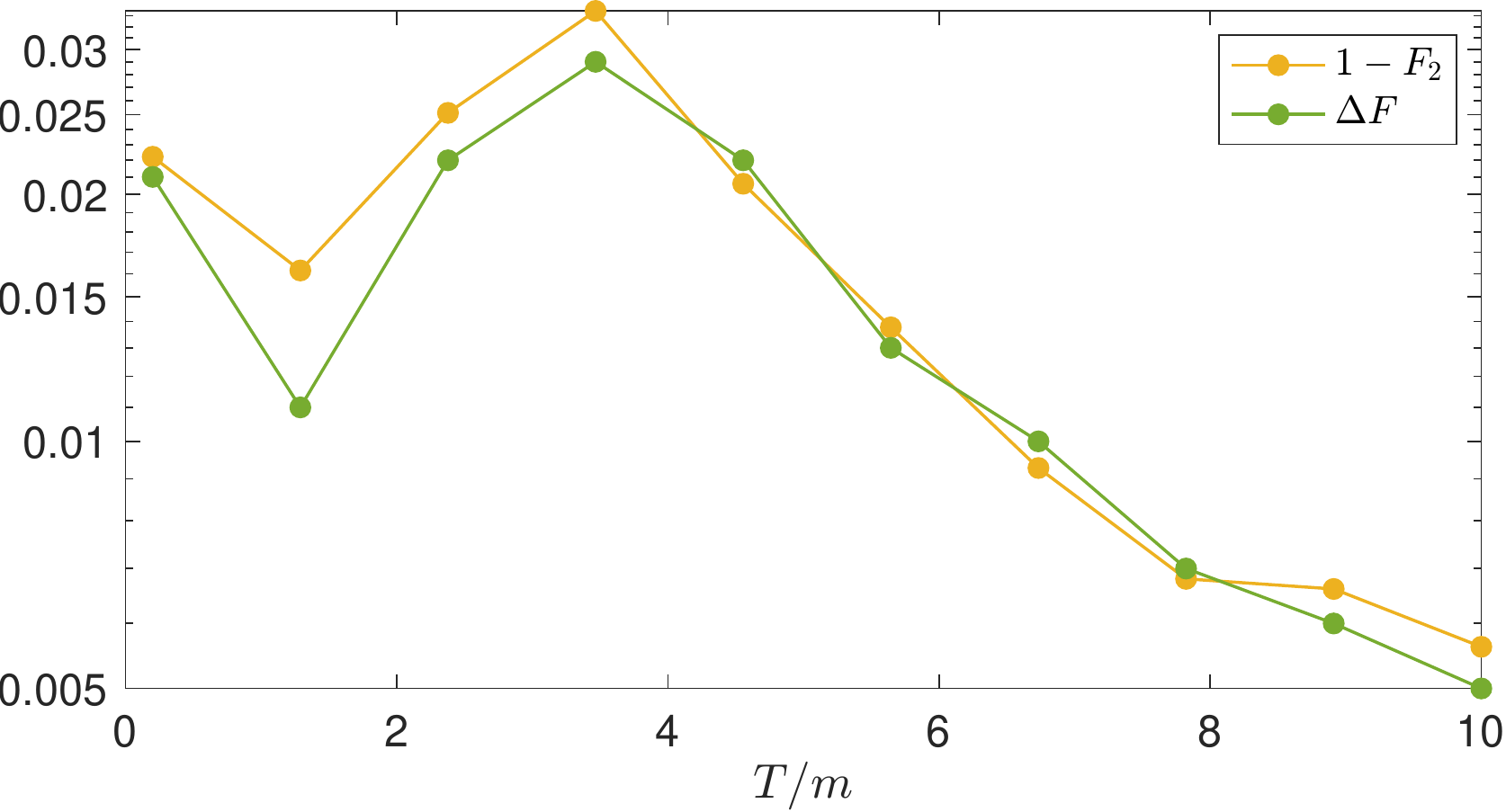} 
\includegraphics[scale=.5]{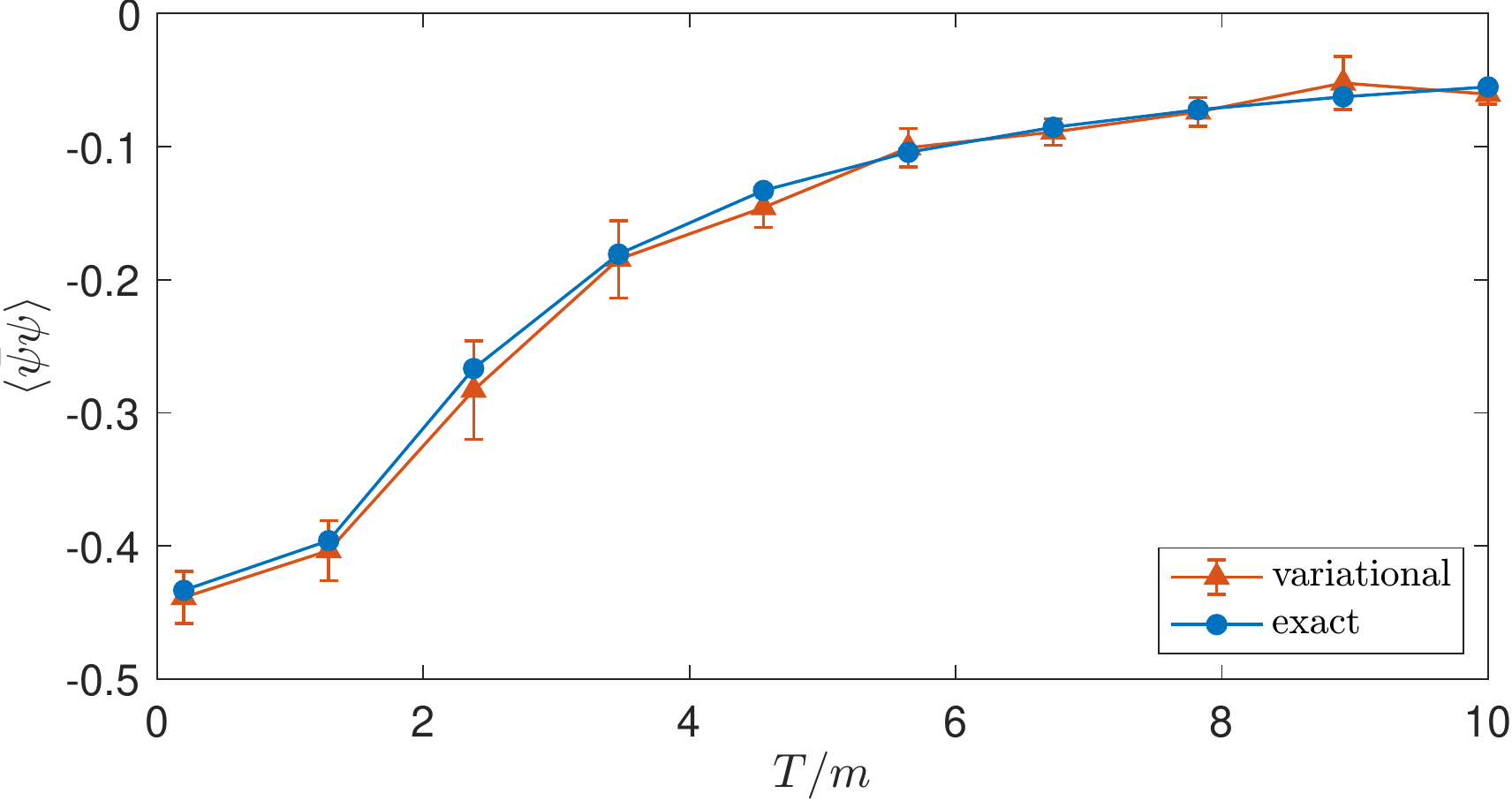}
\caption{\label{fig:horn_results}\emph{(Left)} The fidelity of the approximation to the Gibbs state produced by VQT on the classical simulator and $\Delta F$ as a function of $T/m$ at $\mu=0$ on a system with $N=4$ matter sites, corresponding to $n_q = 7$ qubits in the gauge-redundant formulation. \emph{(Right)} The chiral condensate $\langle\bar\psi\psi\rangle$ calculated on the same data sets where for clarity the exact result has also been included.}
\end{figure*}
\section{Implementation and Results}
The $\mathbb{Z}_2$ lattice gauge theory has a chiral symmetry which is spontaneously broken at low $T$. At large temperatures as well as at large values of the chemical potential, this symmetry is restored. To benchmark our approximation to the Gibbs state, we calculate the chiral condensate, $\vev{\bpsi \psi}$, which is sensitive to restoring chiral symmetry. As this is a local, time-independent observable, one would like a second, time-dependent observable that measures correlations in the system. We chose to study the unequal-time density-density correlator, which provides information about charge dynamics in the system.
\subsection{Gauge redundant formulation}
In this formulation of the theory, gauge invariance has to be considered explicitly. Apart from the gauge-invariant initial state (c.f.~Fig.~\ref{fig:initial_state}), this not only applies to the variational circuits $\mathrm{VQC_{1/2}}$ but also to the intermediate measurement. A schematic depiction of the circuit illustrating the above-mentioned components is shown in Fig.~\ref{fig:vqt_sketch}. For $\mathrm{VQC_{1/2}}$, a gauge-invariant gate set can be constructed by following the general considerations laid out in~\cite{Mazzola2021}. This consists of simply taking the Pauli strings in our Hamiltonian Eq.(\ref{eq:Z2horn}) as elementary building blocks. Due to the simplicity of the model, the gate set will hence decompose into parameter-dependent single qubit $R_x$ and $R_z$ gates, together with a 3-qubit gate $\sim e^{i\sigma^x Z \sigma^x + i\sigma^yZ\sigma^y}$ stemming from the hopping term in Eq.(\ref{eq:Z2horn}). Following the procedure presented in~\cite{Murairi2022}, the latter can be easily decomposed into single-qubit and entangling gates. In Fig.~\ref{fig:Horn_varcirc}, one observes the resulting circuit layer consisting of a sublayer of single-qubit gates along with a sublayer of the decomposed multi-qubit gate for a three-qubit system where the $\Delta_i$ represents generic variational parameters. Several such layers then yield the $\mathrm{VQC}_i$, with parameter sets $\boldsymbol\theta$ and $\boldsymbol\phi$ for $i =1,2$, respectively (Eq.(\ref{eq:rho_var})). With the composition of the variational quantum circuits at hand, we still need to define the figures of merit quantifying the quality of the optimization step depicted in the classical block of Fig.~\ref{fig:vqt_sketch}. In addition to the relative difference in free energy, $\Delta F = F_{\rho_{var}}/F_{\rho_{\text{Gibbs}}}-1$, we take the fidelity~\cite{MichaelA.Nielsen2010}
\beq \label{eq:Fidelity}
F_2(\rho_1,\rho_2) = \left( \Tr \sqrt{\sqrt{\rho_1} \rho_2 \sqrt{\rho_2}} \right)^2, 
\eeq 
as a distance measure between two quantum states represented by  $\rho_1$ and $\rho_2$, respectively. In Fig.~\ref{fig:horn_results}, one sees the results of the classically simulated variational optimization process for our gauge-redundant theory for $N = 4$ matter sites corresponding to $n_q = 7$ qubits using open boundary conditions with a two-layer ansatz in the $\mathrm{VQC}_i$. Simulating at $a\epsilon = a m = 0.5, a\mu = 0$ and $n_s = 10^4$ for varying $T/m$, chiral symmetry is explicitly broken in our model. On the left, we plot the performance measures defined above, where each point corresponds to the minimum in free energy over up to $10^2$ samples, resulting in optimal parameter sets $(\boldsymbol\theta^*,\boldsymbol\phi^*)$. The quality of the approximation by the variational ansatz decreases slightly in the crossover region until, with increasing $T/m$, the entropic term in Eq.(\ref{eq:nonzeroTvarpriniciple}) becomes dominant, yielding a high-fidelity solution. On the right, the chiral condensate $\langle\bar\psi\psi\rangle$ evaluated using $\rho_2(\boldsymbol\theta^*,\boldsymbol\phi^*)$ is shown as a function of $T/m$. The displayed errors were obtained from the variation of the observable over solutions corresponding to the $20$th percentile of the free energy distribution for each set of samples.
\begin{figure}
\scalebox{1.0}{
\Qcircuit @C=1.0em @R=0.2em @!R { \\
	 	\nghost{{q}_{0} :  } & \lstick{{q}_{0} :  } & \qw & \qw & \qw & \targ & \gate{\mathrm{R_Z}\,(\mathrm{{\ensuremath{\Delta}}})} & \targ & \qw & \qw & \qw \barrier[0em]{2} & \qw & \qw & \qw\\
	 	\nghost{{q}_{1} :  } & \lstick{{q}_{1} :  } & \gate{\mathrm{S^\dagger}} & \gate{\mathrm{H}} & \targ & \ctrl{-1} & \qw & \ctrl{-1} & \targ & \gate{\mathrm{H}} & \gate{\mathrm{S}} & \qw & \qw & \qw\\
	 	\nghost{{q}_{2} :  } & \lstick{{q}_{2} :  } & \qw & \qw & \ctrl{-1} & \qw & \qw & \qw & \ctrl{-1} & \qw & \qw & \qw & \qw & \qw\\
\\ }}
\caption{\label{fig:Zohar_varcirc} The decomposed three-qubit gate corresponding to the three boson coupling in Eq.(\ref{eq:H_Zohar_Z2_static_background_fixed}).}
\end{figure}
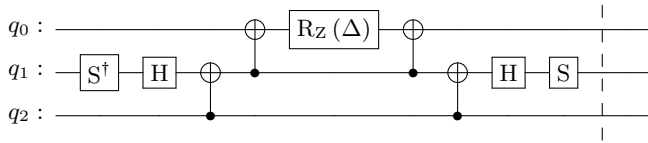
\subsection{Resource efficient formulation}
Turning now to the Hamiltonian Eqs. (\ref{eq:H_Zohar_Z2_static_background_fixed}) with matter degrees of freedom eliminated and gauge invariance \emph{built-in}, the choice of variational ansatz circuits for the $\mathrm{VQC}_i$ is in principle unrestricted, leaving us with the freedom to use, e.g. hardware efficient circuits such as EfficientSU2 from qiskit's standard circuit library \cite{Qiskit}. As for the set of physical gates, by inspecting the Hamiltonian one sees that the single and multi-qubit terms can be employed as elementary ansätze. As an exaample, the decomposed three-qubit term is depicted in Fig.~\ref{fig:Zohar_varcirc}. An alternation of a single-qubit sublayer containing $R_x$ and $R_y$ gates, followed by a staggered multi-qubit sublayer, will hence serve as our variational ansatz for both $\mathrm{VQC}_{1/2}$.
\subsubsection{Observables at $T,\mu > 0$}
\begin{figure*}[t]
\includegraphics[scale=.5]{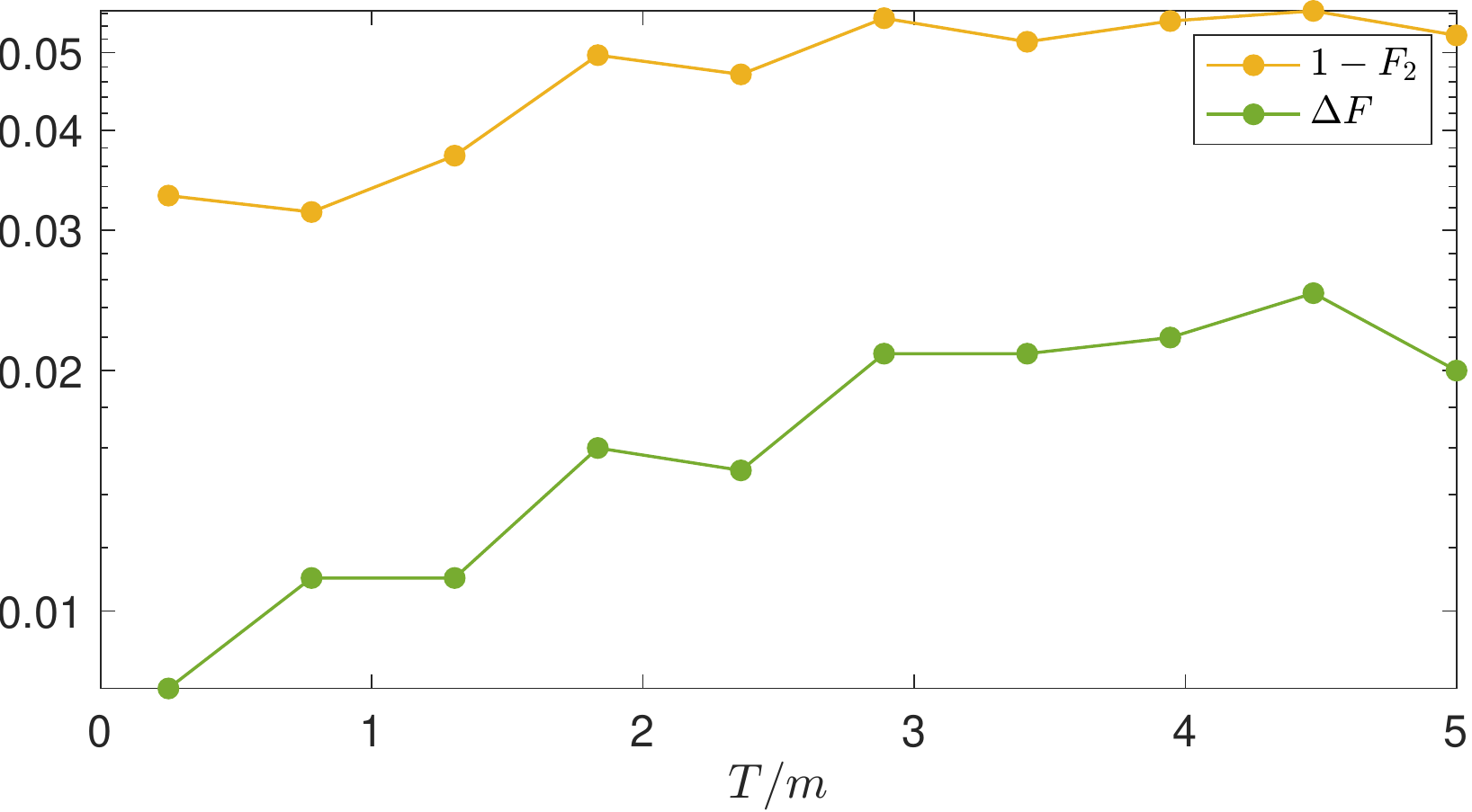} 
\includegraphics[scale=.5]{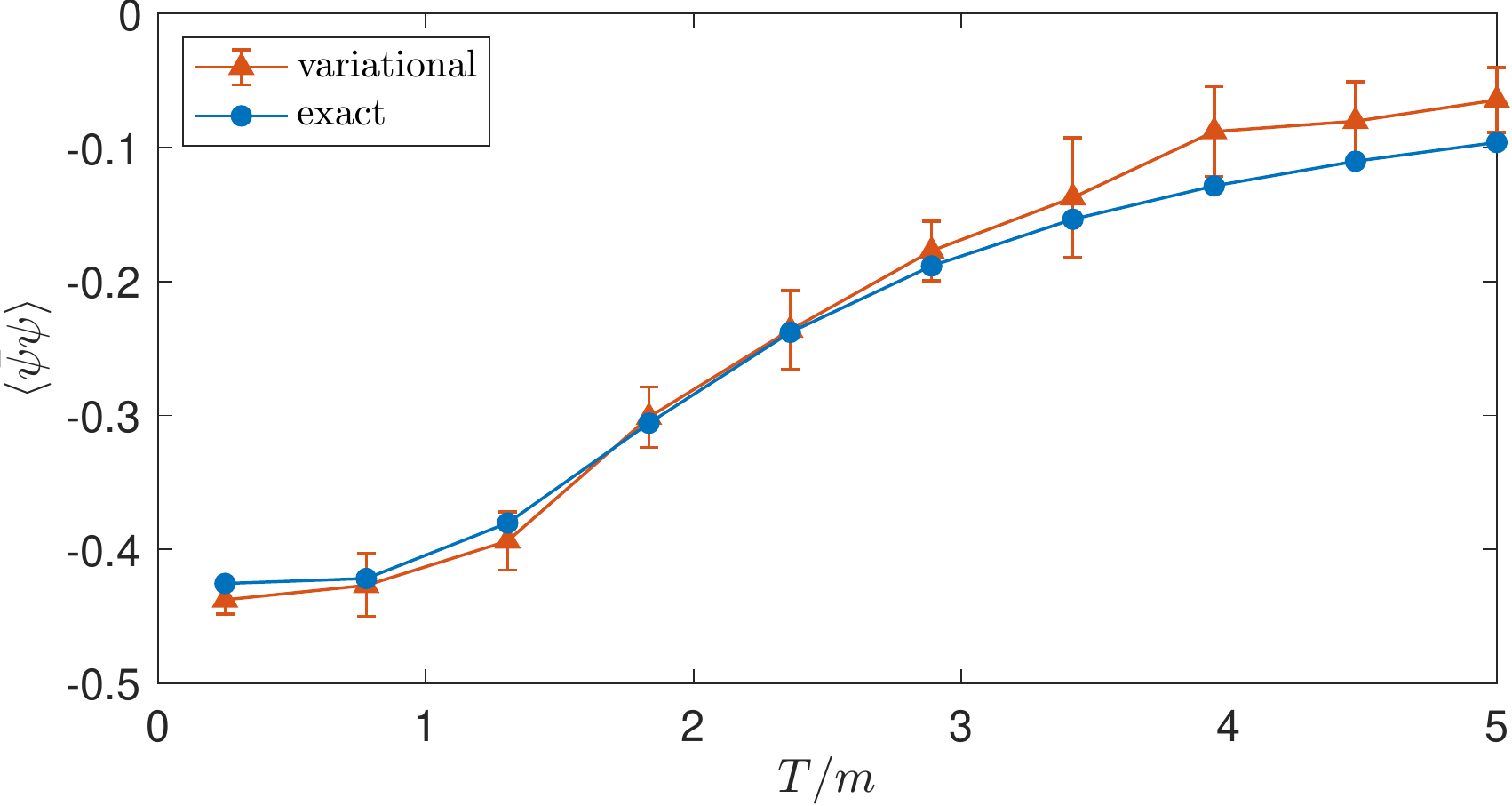}
\caption{\label{fig:Zohar_results_1}\emph{(Left)} The fidelity of the approximation to the Gibbs state produced by VQT on the classical simulator and $\Delta f$ as a function of $T/m$ for $\mu = 0$ on a system with $N=8$ matter sites, corresponding to $n_q = 7$ qubits in the resource-efficient formulation. \emph{(Right)} The chiral condensate $\langle\bar\psi\psi\rangle$ calculated on the same data sets.}
\end{figure*}

\begin{figure*}[t]
\includegraphics[scale=.5]{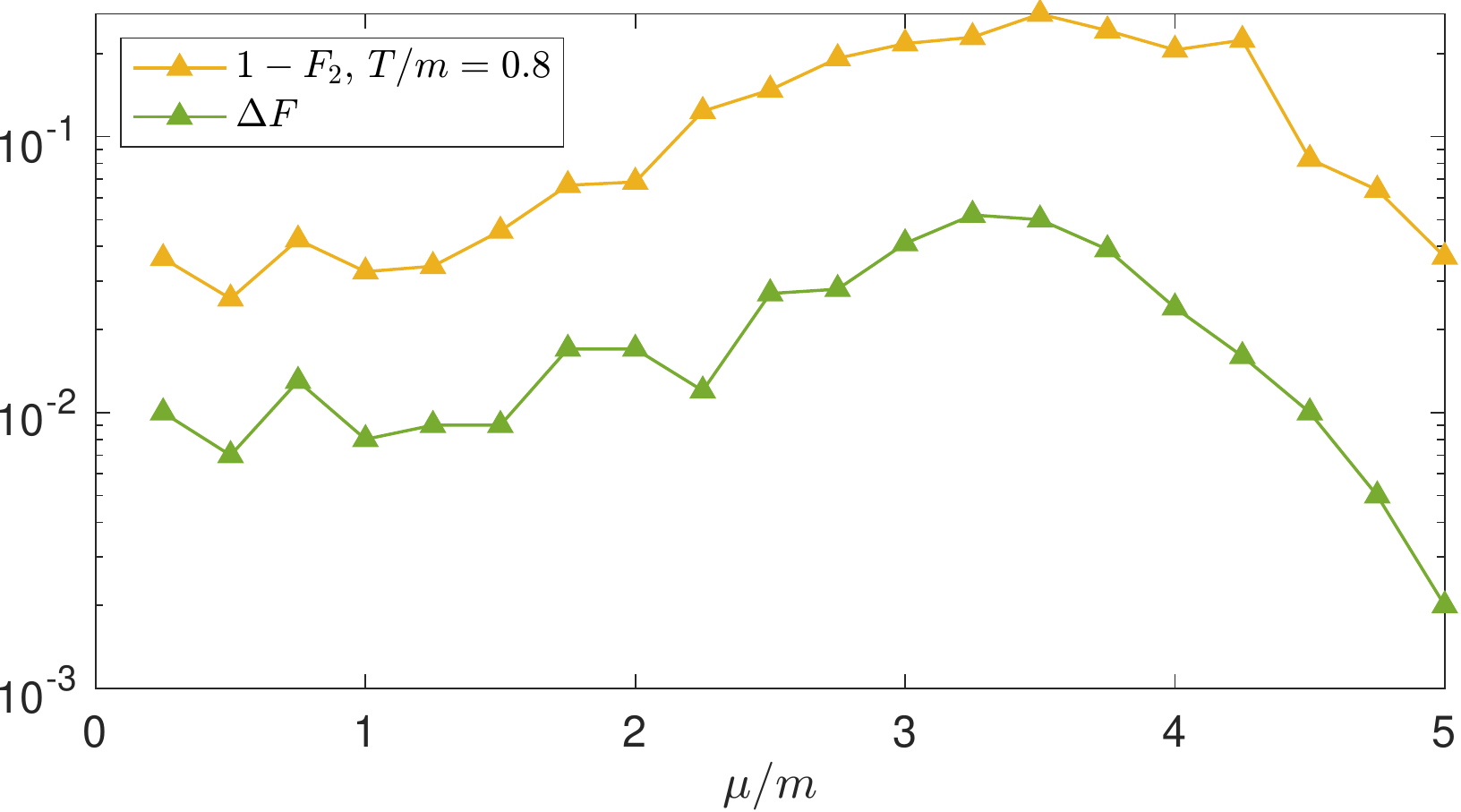} 
\includegraphics[scale=.5]{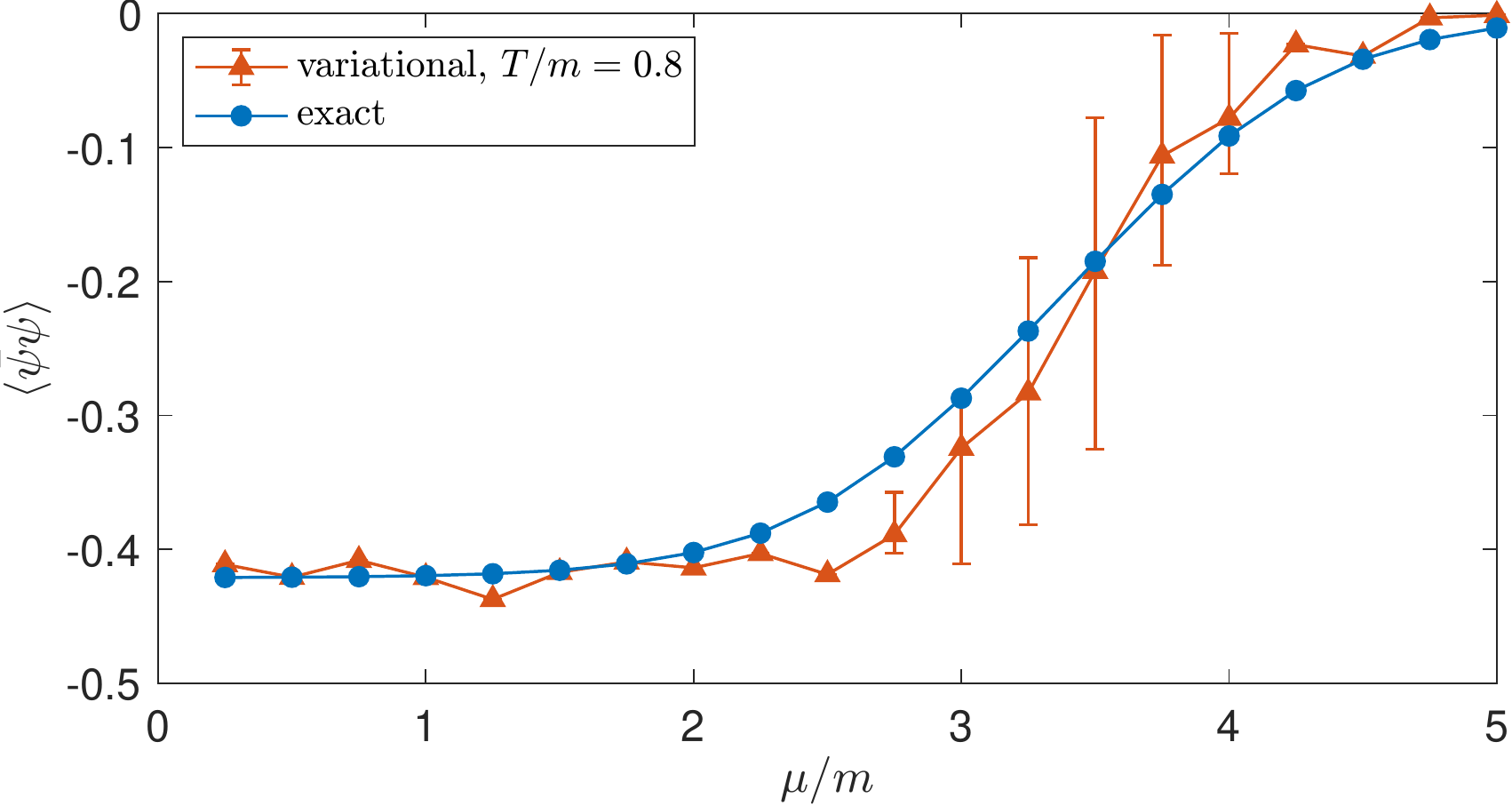}
\caption{\label{fig:Zohar_results_3}\emph{(Left)} The fidelity of the approximation to the Gibbs state produced by VQT on the classical simulator and $\Delta f$ as a function of $\mu/m$ for $T/m = 0.8$ on a system with $N=8$ matter sites, corresponding to $n_q = 7$ qubits in the resource-efficient formulation.\emph{(Right)} The chiral condensate $\langle\bar\psi\psi\rangle$ calculated on the same data sets.}
\end{figure*}

\begin{figure*}[t]
\includegraphics[scale=.5]{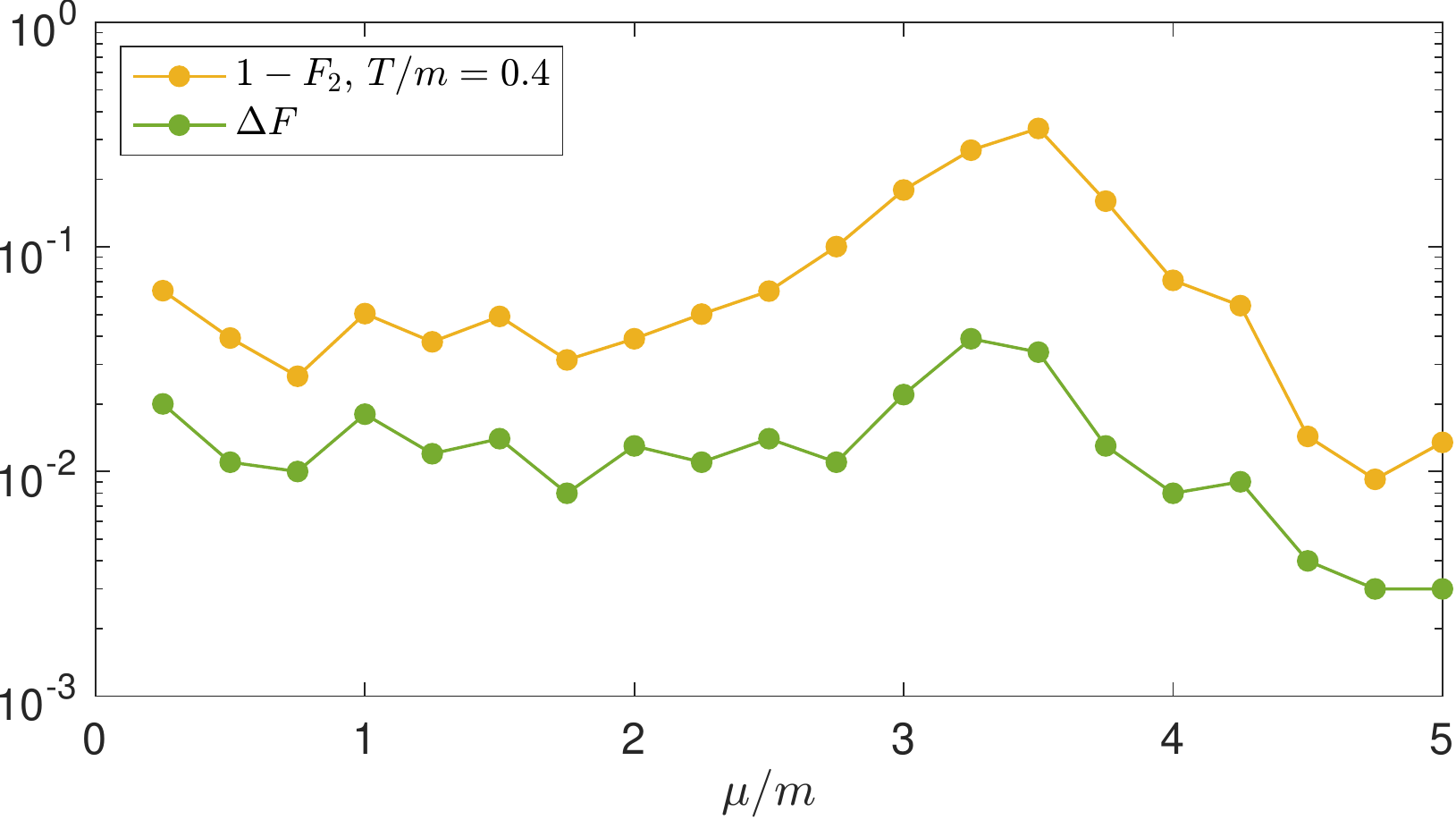} 
\includegraphics[scale=.5]{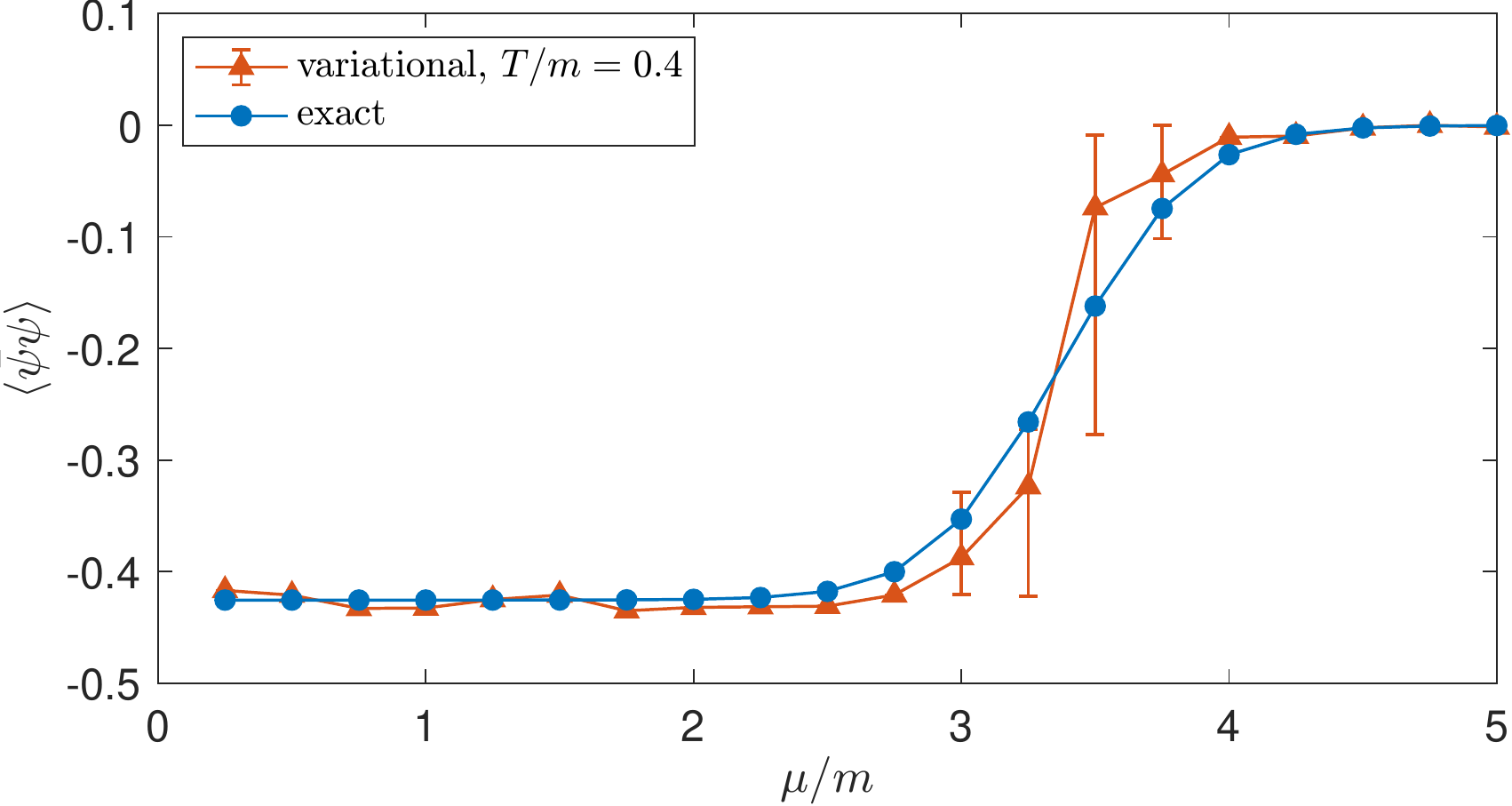}
\caption{\label{fig:Zohar_results_2}\emph{(Left)} The fidelity of the approximation to the Gibbs state produced by VQT on the classical simulator and $\Delta f$ as a function of $\mu/m$ for $T/m = 0.4$ on a system with $N=8$ matter sites, corresponding to $n_q = 7$ qubits in the resource-efficient formulation. \emph{(Right)} The chiral condensate $\langle\bar\psi\psi\rangle$ calculated on the same data sets.}
\end{figure*}

As was done in the gauge-redundant formulation, we quantitatively investigate the quality of our variational approximation to the Gibbs state and calculate the chiral condensate. This has been done as a function of temperature $T/m$ and varying chemical potential $\mu/m$ at $\epsilon/m = 0.5$ with $n_s = 10^4$. Owing to the resource efficiency of the formulation, $n_q = 7$ classically simulated qubits now correspond to twice the system size ($N = 8$) when compared to the gauge-redundant formulation. While simulations for varying $T/m$ across the cross-over region, Fig.~\ref{fig:Zohar_results_1}, show good agreement between exact and simulated results, entering the transition region for fixed temperature and increasing $\mu/m$, we see that our algorithm has difficulty to match the exact result. One notes that as we lower the temperature, the transition becomes more and more discontinuous. In Fig.~\ref{fig:Zohar_results_2} and Fig~\ref{fig:Zohar_results_3}, this observation is confirmed by our estimation of the statistical error which measures the variation of $\langle\bar\psi\psi\rangle$ over the top $20$th percentile of variational solutions. We expect an increase in performance can be achieved here by changing from the gradient-free optimization algorithm (COBYLA) used in this study to gradient-based methods. This is an approach which we have not pursued further.

\subsubsection{Thermal Unequal-Time Correlators}
\label{sect:therm_corr}
In addition to the study of quantum systems at finite fermion density, quantum simulation in the NISQ era and beyond will allow the investigation of real-time observables. These include thermal unequal-time correlation functions. In general, an unequal-time correlator of two operators takes the form
\beq \label{eq:time_dependent_correlator}
C_{AB}(t) \equiv \vev{e^{iHt}Ae^{-iHt}B} = \vev{A(t)B},
\eeq 
where $H$ is the Hamiltonian of the system and $A(t)$ is the operator in the Heisenberg picture. In this study, the expectation value in (\ref{eq:time_dependent_correlator}) will be taken with respect to our variational density matrix, $\rho_2(\boldsymbol\theta^*,\boldsymbol\phi^*)$.

The question of how one measures such a correlation function naturally arises when studying lattice gauge theories. Fortunately, a general procedure has already been devised to measure the correlation function of two arbitrary unitary operators $A$ and $B$ on a universal quantum computer \cite{Ortiz2000,PhysRevA.65.042323}. This procedure is commonly referred to as Ramsey interferometry and only involves adding a single ancillary qubit entangled with our system's original $n_q$ qubits. The time evolution can be performed with a Trotterization of the time evolution operator whose circuit depth increases linearly with the number of steps. A variational approach also exists to obtain the time evolution of an arbitrary quantum state with a fixed circuit depth \cite{Irmejs2022}. In this work, a simple Trotterization has been used.

\begin{figure*}[t]
\includegraphics[scale=.51]{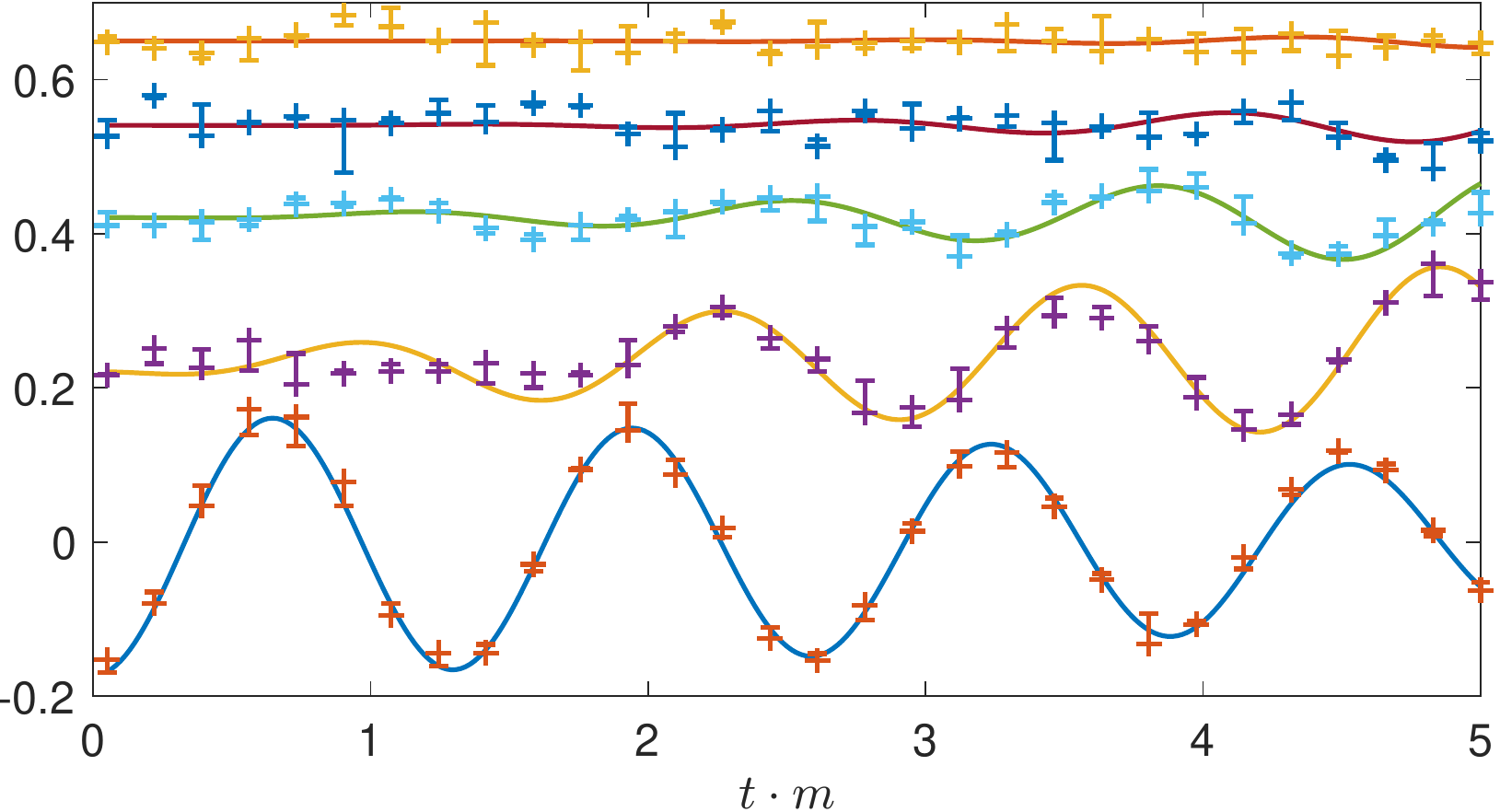}
\includegraphics[scale=.52]{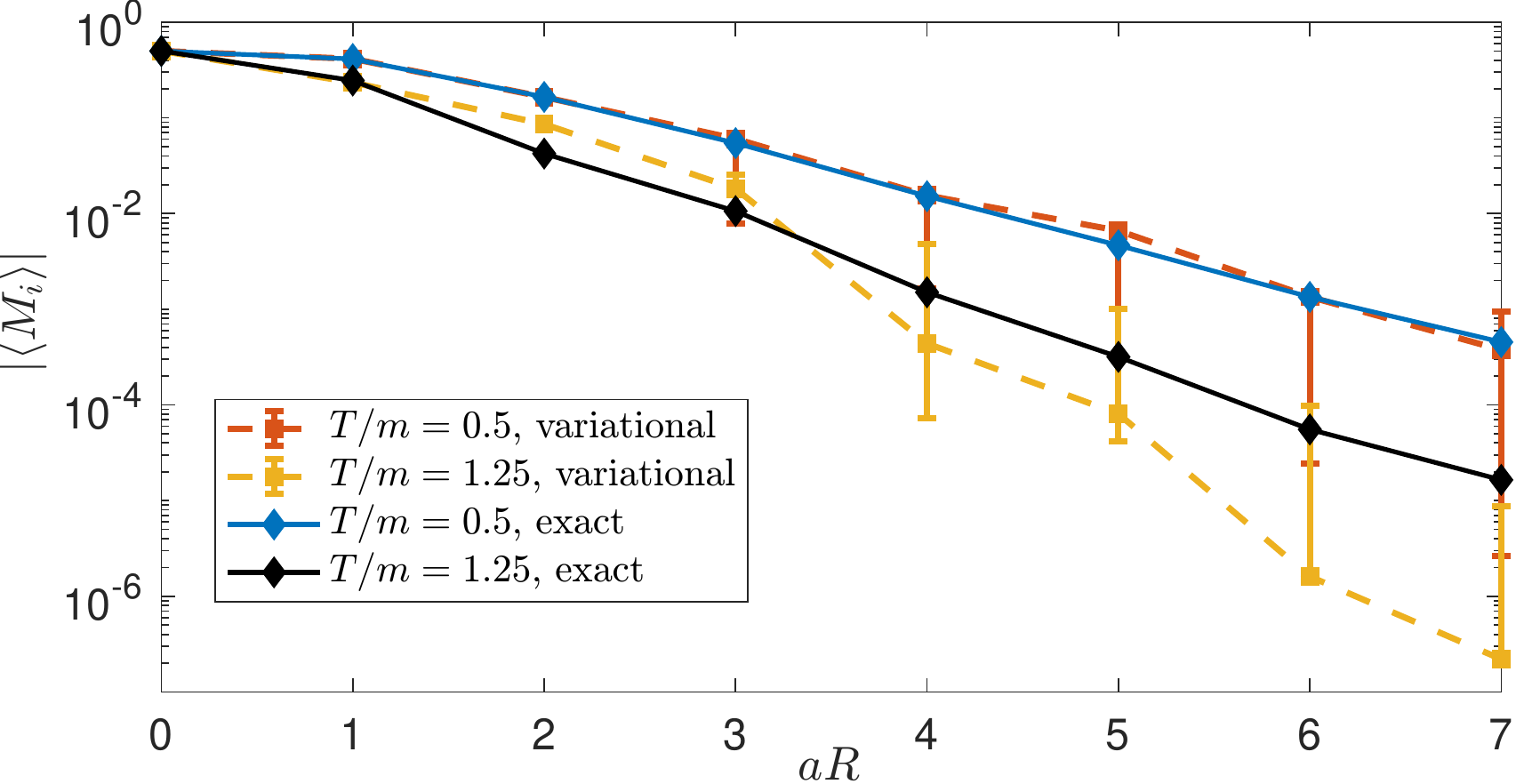}
\caption{\label{fig:n0nit_corr} (Left) The unequal-time thermal correlation function $\langle O_0(0) O_i(t)\rangle$ for varying distance $i = 1,\ldots,5$ (shifted in magnitude for displaying purposes) at $T/m = 1$ for $N = 8$, where $O_i = \varepsilon_iZ_{i-1}\otimes Z_{i}$, i.e. the relevant term of the number operator. The solid lines are the exact solutions while the crosses represent the values calculated with the Trotterized time-evolution operator ($a\delta_t = 0.2$) for the variatonal Gibbs state $\rho_{var}$ produced with infinite statistics. (Right) The absolute value of the meson string operator as a function of the string length (lattice units). The results for two temperatures are plotted using both the variational Gibbs state and the exact density matrix. One notices that at larger temperatures the current variational ansätze give states which deviate from the exact result.}
\end{figure*}

In Fig.~\ref{fig:n0nit_corr} we show the unequal-time thermal density-density correlation function $\langle O_0(0) O_i(t)\rangle$ at $T/m = 1$ for $N = 8$, where $O_i \equiv \varepsilon_iZ_{i-1}\otimes Z_{i}$. The different data sets represent different spatial separations $i = 1,\ldots,5$ of the density operator on the lattice plotted as a function of time with a remarkable agreement between simulated and exact results for the time range displayed. In these classical simulations of our variational quantum circuit, both $\mathrm{VQC}_1$ and $\mathrm{VQC}_2$ contain two layers of alternating single and multi-qubit sublayers. For simplicity's sake, we have assumed infinite statistics in the computation of $\rho_2(\boldsymbol\theta^*,\boldsymbol\phi^*)$, hence excluding shot noise. One is ultimately interested in obtaining transport coefficients from real-time dynamics. As the latter involves time integrals of unequal-time correlators, it remains to be seen if such simulations yield sufficient accuracy.

\subsubsection{$\bar q q$ meson screening}
It is useful to check that our algorithm correctly reproduces physical observables, which 
are more easily accessible by standard Eulcidean methods. As an example, we consider the thermal expectation
value of a gauge-invariant $\bar q q$ meson operator, i.e.~a quark anti-quark pair connected by 
a flux tube~\cite{ZoharZ22022, Zohar2019},
\beq \label{eq:MesonStringOp}
M_i \equiv \psi^\dagger_0 \left( \prod^{i-1}_{k=0} X_k \right) \psi_i.
\eeq 
We have written the above operator in the gauge-redundant version for simplicity but have used the corresponding resource-efficient expression in our calculations.
As demonstrated recently with the help of DMRG~\cite{Borla2019}, at $T=0$, this equal-time two-point function shows an exponential decay for any $\epsilon > 0$. The exponents of a spatial decay are screening masses, which in this case give the energy of the flux tube and
its excitations, whose linear increase with separation indicates confinement.  
We expect this behaviour to persist for small temperatures until the onset of string breaking, when the flux
gets increasingly screened and its energy no longer grows with separation. 

Using the same variational setup as described in Sect.\ref{sect:therm_corr}, Fig.~\ref{fig:n0nit_corr} (right) shows $|\langle M_i\rangle|$  at $T/m = 0.5$ and $T/m = 1.25$. Whereas the low-temperature correlator exhibits marked similarities to the zero-temperature decay (c.f.~\cite{Borla2019}), which is accurately captured by the variational Gibbs state, at higher temperatures, the algorithm fails to reproduce the exact correlator accurately, possibly due to the low complexity of the ansatz circuits in our simulations (two layers of variational gates for the $\mathrm{VQC}_i$). While this precludes a detailed picture of the long distance behavior with increasing temperature, our variational calculation does exhibit two qualitative features known from other techniques: i) the screening masses generally grow with temperature as expected (faster decay); ii) the difference between high and low temperature is small at short distances, but pronounced at large distances, because of the flux screening. 

\subsubsection{Resources and Entropy Estimation}
We now turn to the question of resources in runtime and classical memory. Considering the limited system sizes currently studied in the NISQ era, estimating the von Neumann entropy $S = -\sum_{i} p_i \log{p_i}$ from the intermediate measurements via  $p_i \approx n_i/n_s$ is a viable approach. Applying the algorithm with the information flow depicted in Fig.~\ref{fig:vqt_sketch} for realistic system sizes will necessarily involve a certain level of approximation. This stems from the fact that the number of $p_i$'s grows with the dimension of the Hilbert space, $2^{n_q}$. One possible way proposed in~\cite{Selisko2022}, is the partitioning of the full $n_q$ qubit system into $n_q/n_{ss}$ independent subsystems of size $n_{ss}$ in the entropic part of the algorithm ($\mathrm{VQC}_1$) that prepares the latent distribution $\rho_1(\boldsymbol\theta)$. In the version of the algorithm which employs a classical estimation of the entropy~\cite{Verdon2019} this corresponds to a particular choice of the latent distribution known as factorized latent space models. It remains to be seen if such an approximation is a valid approach for gauge theories where thermalization and subsystem entanglement are non-trivially connected~\cite{Mueller2022}.

\begin{figure*}[t]
\includegraphics[scale=.6]{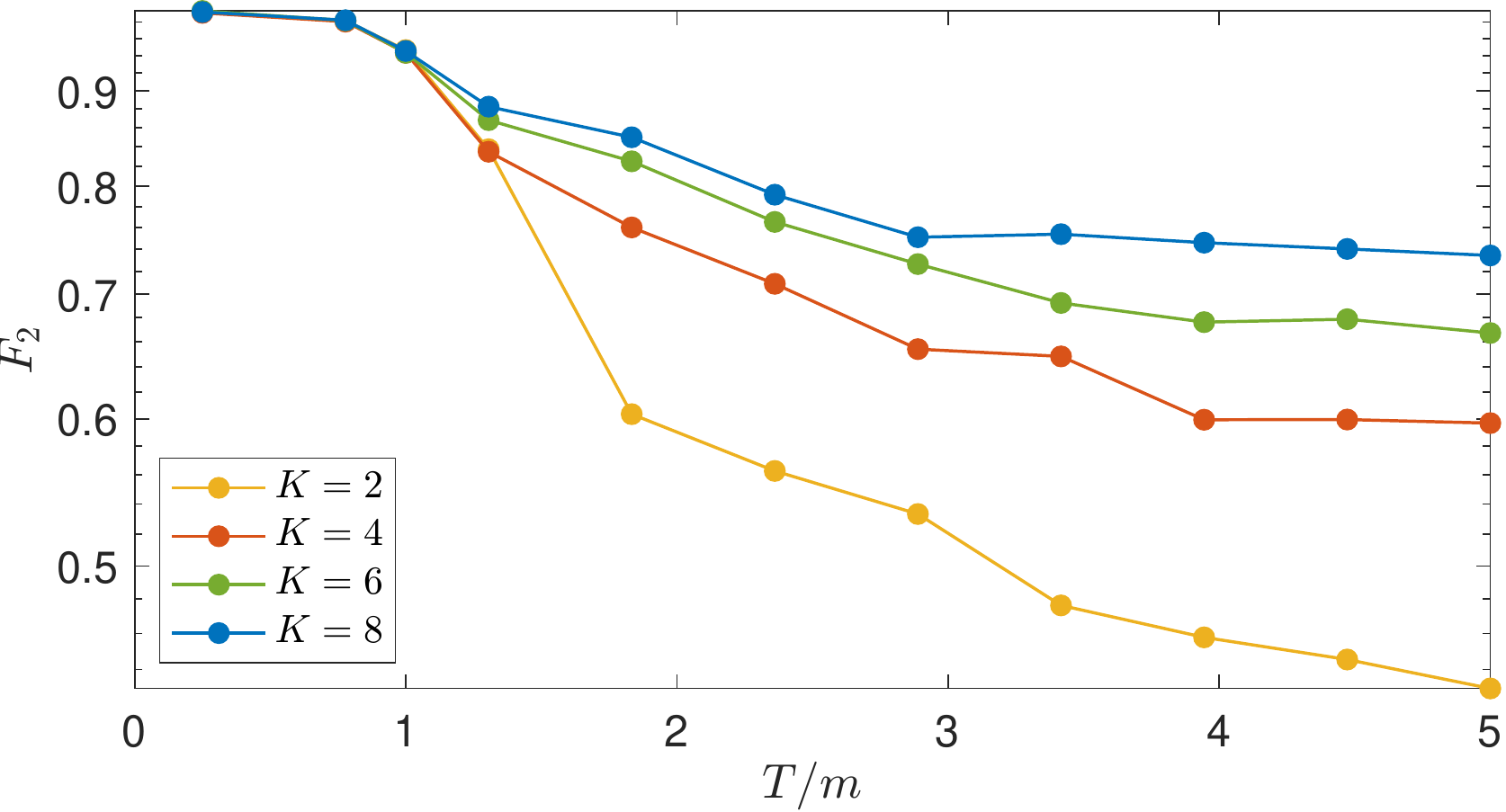} 
\caption{\label{fig:entropy_taylor}The fidelity of the Gibbs state produced using the Taylor-series approximated von Neumann entropy as a function of $T/m$ at $a\mu = 0$ for various truncation orders $K$ on a system with $N = 8$ matter sites. One sees that for low temperatures this approximation works well.}
\end{figure*}

There are, however, alternative methods that can be used to estimate the entropy efficiently. For example, it turns out that an evaluation of the Taylor series approximation to the entropy can be useful~\cite{Wang2021}. This method requires the addition of $n_q$ ancillary bits to perform a qubit-efficient, repeated SWAP-test evaluation of the trace of various powers of the latent density matrix, $\mathrm{tr}(\rho^n_1(\theta))$. The results of this approach for the simple gauge theory used in this study are displayed in Fig.~\ref{fig:entropy_taylor}. For temperatures up to $T/m\sim 1$, only a few orders are needed to give reasonable values of the fidelity.

\label{subsect:resources}

\section{Conclusion}

We have explored the use of variational quantum algorithms in producing thermal states. These approaches were applied to $\mathbb{Z}_2$ lattice gauge theory in $1+1$ dimensions using two independent formulations, and several observables of interest were computed. We find this variant of the VQT a suitable algorithm to approximate the thermal states of quantum systems in the NISQ era as estimating the entropy of larger systems will necessitate the introduction of truncations in the estimation. A particular aspect we have left for future work is the performance of intermediate measurements for mixed-state creation on actual quantum devices compared to the usual approach of entangling ancillary systems followed by a terminal measurement.\\
A further possible extension of our work is the formulation of the VQT to study non-Abelian gauge theories at finite temperatures. As the latter will generally come with a gauge-redundant Hilbert space, even after reformulating in a resource-efficient manner, this amounts to respecting gauge invariance or equivalent local constraints throughout the variational quantum circuit.

\acknowledgments
The work is supported by the Deutsche Forschungsgemeinschaft (DFG, German Research Foundation) through the grant CRC-TR 211 ``Strong-interaction matter under extreme conditions''~--~project number 315477589~--~TRR 211 and by the State of Hesse within the Research Cluster ELEMENTS (Project ID 500/10.006). M.S. and M.F. acknowledge support by the Munich Institute for Astro-, Particle and BioPhysics (MIAPbP) which is funded by the Deutsche Forschungsgemeinschaft (DFG, German Research Foundation) under Germany's Excellence Strategy – EXC-2094 – 390783311.

\bibliography{pub}

\begin{thebibliography}{27}%
\makeatletter
\providecommand \@ifxundefined [1]{%
 \@ifx{#1\undefined}
}%
\providecommand \@ifnum [1]{%
 \ifnum #1\expandafter \@firstoftwo
 \else \expandafter \@secondoftwo
 \fi
}%
\providecommand \@ifx [1]{%
 \ifx #1\expandafter \@firstoftwo
 \else \expandafter \@secondoftwo
 \fi
}%
\providecommand \natexlab [1]{#1}%
\providecommand \enquote  [1]{``#1''}%
\providecommand \bibnamefont  [1]{#1}%
\providecommand \bibfnamefont [1]{#1}%
\providecommand \citenamefont [1]{#1}%
\providecommand \href@noop [0]{\@secondoftwo}%
\providecommand \href [0]{\begingroup \@sanitize@url \@href}%
\providecommand \@href[1]{\@@startlink{#1}\@@href}%
\providecommand \@@href[1]{\endgroup#1\@@endlink}%
\providecommand \@sanitize@url [0]{\catcode `\\12\catcode `\$12\catcode
  `\&12\catcode `\#12\catcode `\^12\catcode `\_12\catcode `\%12\relax}%
\providecommand \@@startlink[1]{}%
\providecommand \@@endlink[0]{}%
\providecommand \url  [0]{\begingroup\@sanitize@url \@url }%
\providecommand \@url [1]{\endgroup\@href {#1}{\urlprefix }}%
\providecommand \urlprefix  [0]{URL }%
\providecommand \Eprint [0]{\href }%
\providecommand \doibase [0]{https://doi.org/}%
\providecommand \selectlanguage [0]{\@gobble}%
\providecommand \bibinfo  [0]{\@secondoftwo}%
\providecommand \bibfield  [0]{\@secondoftwo}%
\providecommand \translation [1]{[#1]}%
\providecommand \BibitemOpen [0]{}%
\providecommand \bibitemStop [0]{}%
\providecommand \bibitemNoStop [0]{.\EOS\space}%
\providecommand \EOS [0]{\spacefactor3000\relax}%
\providecommand \BibitemShut  [1]{\csname bibitem#1\endcsname}%
\let\auto@bib@innerbib\@empty
\bibitem [{\citenamefont {Ba\~nuls}\ \emph {et~al.}(2020)\citenamefont
  {Ba\~nuls} \emph {et~al.}}]{Banuls:2019bmf}%
  \BibitemOpen
  \bibfield  {author} {\bibinfo {author} {\bibfnamefont {M.~C.}\ \bibnamefont
  {Ba\~nuls}} \emph {et~al.},\ }\bibfield  {title} {\bibinfo {title}
  {{Simulating Lattice Gauge Theories within Quantum Technologies}},\ }\href
  {https://doi.org/10.1140/epjd/e2020-100571-8} {\bibfield  {journal} {\bibinfo
   {journal} {Eur. Phys. J. D}\ }\textbf {\bibinfo {volume} {74}},\ \bibinfo
  {pages} {165} (\bibinfo {year} {2020})},\ \Eprint
  {https://arxiv.org/abs/1911.00003} {arXiv:1911.00003 [quant-ph]} \BibitemShut
  {NoStop}%
\bibitem [{\citenamefont {Bauer}\ \emph {et~al.}(2023)\citenamefont {Bauer}
  \emph {et~al.}}]{Bauer:2022hpo}%
  \BibitemOpen
  \bibfield  {author} {\bibinfo {author} {\bibfnamefont {C.~W.}\ \bibnamefont
  {Bauer}} \emph {et~al.},\ }\bibfield  {title} {\bibinfo {title} {{Quantum
  Simulation for High-Energy Physics}},\ }\href
  {https://doi.org/10.1103/PRXQuantum.4.027001} {\bibfield  {journal} {\bibinfo
   {journal} {PRX Quantum}\ }\textbf {\bibinfo {volume} {4}},\ \bibinfo {pages}
  {027001} (\bibinfo {year} {2023})},\ \Eprint
  {https://arxiv.org/abs/2204.03381} {arXiv:2204.03381 [quant-ph]} \BibitemShut
  {NoStop}%
\bibitem [{\citenamefont {Wu}\ and\ \citenamefont {Hsieh}(2018)}]{Wu2018}%
  \BibitemOpen
  \bibfield  {author} {\bibinfo {author} {\bibfnamefont {J.}~\bibnamefont
  {Wu}}\ and\ \bibinfo {author} {\bibfnamefont {T.~H.}\ \bibnamefont {Hsieh}},\
  }\bibfield  {title} {\bibinfo {title} {Variational thermal quantum simulation
  via thermofield double states},\ }\href
  {https://doi.org/10.1103/PhysRevLett.123.220502} {\bibfield  {journal}
  {\bibinfo  {journal} {Phys. Rev. Lett. 123, 220502 (2019)}\ }\textbf
  {\bibinfo {volume} {123}},\ \bibinfo {pages} {220502} (\bibinfo {year}
  {2018})},\ \Eprint {https://arxiv.org/abs/1811.11756} {arXiv:1811.11756
  [cond-mat.str-el]} \BibitemShut {NoStop}%
\bibitem [{\citenamefont {Ball}\ and\ \citenamefont {Cohen}(2022)}]{Ball2022}%
  \BibitemOpen
  \bibfield  {author} {\bibinfo {author} {\bibfnamefont {C.}~\bibnamefont
  {Ball}}\ and\ \bibinfo {author} {\bibfnamefont {T.~D.}\ \bibnamefont
  {Cohen}},\ }\bibfield  {title} {\bibinfo {title} {Boltzmann distributions on
  a quantum computer via active cooling}\ }\href
  {https://doi.org/10.48550/ARXIV.2212.06730} {10.48550/ARXIV.2212.06730}
  (\bibinfo {year} {2022}),\ \Eprint {https://arxiv.org/abs/2212.06730}
  {arXiv:2212.06730 [quant-ph]} \BibitemShut {NoStop}%
\bibitem [{\citenamefont {Powers}\ \emph {et~al.}(2021)\citenamefont {Powers},
  \citenamefont {Oftelie}, \citenamefont {Camps},\ and\ \citenamefont
  {de~Jong}}]{Powers2021}%
  \BibitemOpen
  \bibfield  {author} {\bibinfo {author} {\bibfnamefont {C.}~\bibnamefont
  {Powers}}, \bibinfo {author} {\bibfnamefont {L.~B.}\ \bibnamefont {Oftelie}},
  \bibinfo {author} {\bibfnamefont {D.}~\bibnamefont {Camps}},\ and\ \bibinfo
  {author} {\bibfnamefont {W.~A.}\ \bibnamefont {de~Jong}},\ }\bibfield
  {title} {\bibinfo {title} {Exploring finite temperature properties of
  materials with quantum computers},\ }\href@noop {} {\  (\bibinfo {year}
  {2021})},\ \Eprint {https://arxiv.org/abs/2109.01619} {arXiv:2109.01619
  [quant-ph]} \BibitemShut {NoStop}%
\bibitem [{\citenamefont {Davoudi}\ \emph {et~al.}(2022)\citenamefont
  {Davoudi}, \citenamefont {Mueller},\ and\ \citenamefont
  {Powers}}]{Davoudi2022}%
  \BibitemOpen
  \bibfield  {author} {\bibinfo {author} {\bibfnamefont {Z.}~\bibnamefont
  {Davoudi}}, \bibinfo {author} {\bibfnamefont {N.}~\bibnamefont {Mueller}},\
  and\ \bibinfo {author} {\bibfnamefont {C.}~\bibnamefont {Powers}},\
  }\bibfield  {title} {\bibinfo {title} {Toward quantum computing phase
  diagrams of gauge theories with thermal pure quantum states},\ }\href@noop {}
  {\  (\bibinfo {year} {2022})},\ \Eprint {https://arxiv.org/abs/2208.13112}
  {arXiv:2208.13112 [hep-lat]} \BibitemShut {NoStop}%
\bibitem [{\citenamefont {Verdon}\ \emph {et~al.}(2019)\citenamefont {Verdon},
  \citenamefont {Marks}, \citenamefont {Nanda}, \citenamefont {Leichenauer},\
  and\ \citenamefont {Hidary}}]{Verdon2019}%
  \BibitemOpen
  \bibfield  {author} {\bibinfo {author} {\bibfnamefont {G.}~\bibnamefont
  {Verdon}}, \bibinfo {author} {\bibfnamefont {J.}~\bibnamefont {Marks}},
  \bibinfo {author} {\bibfnamefont {S.}~\bibnamefont {Nanda}}, \bibinfo
  {author} {\bibfnamefont {S.}~\bibnamefont {Leichenauer}},\ and\ \bibinfo
  {author} {\bibfnamefont {J.}~\bibnamefont {Hidary}},\ }\bibfield  {title}
  {\bibinfo {title} {Quantum hamiltonian-based models and the variational
  quantum thermalizer algorithm},\ }\href@noop {} {\  (\bibinfo {year}
  {2019})},\ \Eprint {https://arxiv.org/abs/1910.02071} {arXiv:1910.02071
  [quant-ph]} \BibitemShut {NoStop}%
\bibitem [{\citenamefont {Peruzzo}\ \emph {et~al.}(2014)\citenamefont
  {Peruzzo}, \citenamefont {McClean}, \citenamefont {Shadbolt}, \citenamefont
  {Yung}, \citenamefont {Zhou}, \citenamefont {Love}, \citenamefont
  {Aspuru-Guzik},\ and\ \citenamefont {O'Brien}}]{Peruzzo2014}%
  \BibitemOpen
  \bibfield  {author} {\bibinfo {author} {\bibfnamefont {A.}~\bibnamefont
  {Peruzzo}}, \bibinfo {author} {\bibfnamefont {J.}~\bibnamefont {McClean}},
  \bibinfo {author} {\bibfnamefont {P.}~\bibnamefont {Shadbolt}}, \bibinfo
  {author} {\bibfnamefont {M.-H.}\ \bibnamefont {Yung}}, \bibinfo {author}
  {\bibfnamefont {X.-Q.}\ \bibnamefont {Zhou}}, \bibinfo {author}
  {\bibfnamefont {P.~J.}\ \bibnamefont {Love}}, \bibinfo {author}
  {\bibfnamefont {A.}~\bibnamefont {Aspuru-Guzik}},\ and\ \bibinfo {author}
  {\bibfnamefont {J.~L.}\ \bibnamefont {O'Brien}},\ }\bibfield  {title}
  {\bibinfo {title} {A variational eigenvalue solver on a photonic quantum
  processor},\ }\bibfield  {journal} {\bibinfo  {journal} {Nature
  Communications}\ }\textbf {\bibinfo {volume} {5}},\ \href
  {https://doi.org/10.1038/ncomms5213} {10.1038/ncomms5213} (\bibinfo {year}
  {2014})\BibitemShut {NoStop}%
\bibitem [{\citenamefont {McClean}\ \emph {et~al.}(2015)\citenamefont
  {McClean}, \citenamefont {Romero}, \citenamefont {Babbush},\ and\
  \citenamefont {Aspuru-Guzik}}]{McClean2015}%
  \BibitemOpen
  \bibfield  {author} {\bibinfo {author} {\bibfnamefont {J.~R.}\ \bibnamefont
  {McClean}}, \bibinfo {author} {\bibfnamefont {J.}~\bibnamefont {Romero}},
  \bibinfo {author} {\bibfnamefont {R.}~\bibnamefont {Babbush}},\ and\ \bibinfo
  {author} {\bibfnamefont {A.}~\bibnamefont {Aspuru-Guzik}},\ }\bibfield
  {title} {\bibinfo {title} {The theory of variational hybrid quantum-classical
  algorithms},\ }\bibfield  {journal} {\bibinfo  {journal} {New J. Phys. 18
  (2016) 023023}\ }\href {https://doi.org/10.1088/1367-2630/18/2/023023}
  {10.1088/1367-2630/18/2/023023} (\bibinfo {year} {2015}),\ \Eprint
  {https://arxiv.org/abs/1509.04279} {arXiv:1509.04279 [quant-ph]} \BibitemShut
  {NoStop}%
\bibitem [{\citenamefont {Foldager}\ \emph {et~al.}(2021)\citenamefont
  {Foldager}, \citenamefont {Pesah},\ and\ \citenamefont
  {Hansen}}]{Foldager2021}%
  \BibitemOpen
  \bibfield  {author} {\bibinfo {author} {\bibfnamefont {J.}~\bibnamefont
  {Foldager}}, \bibinfo {author} {\bibfnamefont {A.}~\bibnamefont {Pesah}},\
  and\ \bibinfo {author} {\bibfnamefont {L.~K.}\ \bibnamefont {Hansen}},\
  }\bibfield  {title} {\bibinfo {title} {Noise-assisted variational quantum
  thermalization},\ }\href@noop {} {\  (\bibinfo {year} {2021})},\ \Eprint
  {https://arxiv.org/abs/2111.03935} {arXiv:2111.03935 [quant-ph]} \BibitemShut
  {NoStop}%
\bibitem [{\citenamefont {Selisko}\ \emph {et~al.}(2022)\citenamefont
  {Selisko}, \citenamefont {Amsler}, \citenamefont {Hammerschmidt},
  \citenamefont {Drautz},\ and\ \citenamefont {Eckl}}]{Selisko2022}%
  \BibitemOpen
  \bibfield  {author} {\bibinfo {author} {\bibfnamefont {J.}~\bibnamefont
  {Selisko}}, \bibinfo {author} {\bibfnamefont {M.}~\bibnamefont {Amsler}},
  \bibinfo {author} {\bibfnamefont {T.}~\bibnamefont {Hammerschmidt}}, \bibinfo
  {author} {\bibfnamefont {R.}~\bibnamefont {Drautz}},\ and\ \bibinfo {author}
  {\bibfnamefont {T.}~\bibnamefont {Eckl}},\ }\bibfield  {title} {\bibinfo
  {title} {Extending the variational quantum eigensolver to finite
  temperatures},\ }\href@noop {} {\  (\bibinfo {year} {2022})},\ \Eprint
  {https://arxiv.org/abs/2208.07621} {arXiv:2208.07621 [quant-ph]} \BibitemShut
  {NoStop}%
\bibitem [{\citenamefont {Consiglio}\ \emph {et~al.}(2023)\citenamefont
  {Consiglio}, \citenamefont {Settino}, \citenamefont {Giordano}, \citenamefont
  {Mastroianni}, \citenamefont {Plastina}, \citenamefont {Lorenzo},
  \citenamefont {Maniscalco}, \citenamefont {Goold},\ and\ \citenamefont
  {Apollaro}}]{Consiglio2023}%
  \BibitemOpen
  \bibfield  {author} {\bibinfo {author} {\bibfnamefont {M.}~\bibnamefont
  {Consiglio}}, \bibinfo {author} {\bibfnamefont {J.}~\bibnamefont {Settino}},
  \bibinfo {author} {\bibfnamefont {A.}~\bibnamefont {Giordano}}, \bibinfo
  {author} {\bibfnamefont {C.}~\bibnamefont {Mastroianni}}, \bibinfo {author}
  {\bibfnamefont {F.}~\bibnamefont {Plastina}}, \bibinfo {author}
  {\bibfnamefont {S.}~\bibnamefont {Lorenzo}}, \bibinfo {author} {\bibfnamefont
  {S.}~\bibnamefont {Maniscalco}}, \bibinfo {author} {\bibfnamefont
  {J.}~\bibnamefont {Goold}},\ and\ \bibinfo {author} {\bibfnamefont
  {T.~J.~G.}\ \bibnamefont {Apollaro}},\ }\bibfield  {title} {\bibinfo {title}
  {Variational gibbs state preparation on nisq devices}\ }\href
  {https://doi.org/10.48550/ARXIV.2303.11276} {10.48550/ARXIV.2303.11276}
  (\bibinfo {year} {2023}),\ \Eprint {https://arxiv.org/abs/2303.11276}
  {arXiv:2303.11276 [quant-ph]} \BibitemShut {NoStop}%
\bibitem [{\citenamefont {Horn}\ \emph {et~al.}(1979)\citenamefont {Horn},
  \citenamefont {Weinstein},\ and\ \citenamefont {Yankielowicz}}]{Horn1979}%
  \BibitemOpen
  \bibfield  {author} {\bibinfo {author} {\bibfnamefont {D.}~\bibnamefont
  {Horn}}, \bibinfo {author} {\bibfnamefont {M.}~\bibnamefont {Weinstein}},\
  and\ \bibinfo {author} {\bibfnamefont {S.}~\bibnamefont {Yankielowicz}},\
  }\bibfield  {title} {\bibinfo {title} {Hamiltonian approach to $z(n)$ lattice
  gauge theories},\ }\href {https://doi.org/10.1103/PhysRevD.19.3715}
  {\bibfield  {journal} {\bibinfo  {journal} {Phys. Rev. D}\ }\textbf {\bibinfo
  {volume} {19}},\ \bibinfo {pages} {3715} (\bibinfo {year}
  {1979})}\BibitemShut {NoStop}%
\bibitem [{\citenamefont {Greenberg}\ \emph {et~al.}(2022)\citenamefont
  {Greenberg}, \citenamefont {Pardo}, \citenamefont {Fortinsky},\ and\
  \citenamefont {Zohar}}]{ZoharZ22022}%
  \BibitemOpen
  \bibfield  {author} {\bibinfo {author} {\bibfnamefont {T.}~\bibnamefont
  {Greenberg}}, \bibinfo {author} {\bibfnamefont {G.}~\bibnamefont {Pardo}},
  \bibinfo {author} {\bibfnamefont {A.}~\bibnamefont {Fortinsky}},\ and\
  \bibinfo {author} {\bibfnamefont {E.}~\bibnamefont {Zohar}},\ }\bibfield
  {title} {\bibinfo {title} {Resource-efficient quantum simulation of lattice
  gauge theories in arbitrary dimensions: Solving for gauss' law and fermion
  elimination},\ }\href@noop {} {\  (\bibinfo {year} {2022})},\ \Eprint
  {https://arxiv.org/abs/2206.00685} {arXiv:2206.00685 [quant-ph]} \BibitemShut
  {NoStop}%
\bibitem [{\citenamefont {Irmejs}\ \emph {et~al.}(2022)\citenamefont {Irmejs},
  \citenamefont {Banuls},\ and\ \citenamefont {Cirac}}]{Irmejs2022}%
  \BibitemOpen
  \bibfield  {author} {\bibinfo {author} {\bibfnamefont {R.}~\bibnamefont
  {Irmejs}}, \bibinfo {author} {\bibfnamefont {M.~C.}\ \bibnamefont {Banuls}},\
  and\ \bibinfo {author} {\bibfnamefont {J.~I.}\ \bibnamefont {Cirac}},\
  }\bibfield  {title} {\bibinfo {title} {Quantum simulation of z2 lattice gauge
  theory with minimal requirements},\ }\href@noop {} {\  (\bibinfo {year}
  {2022})},\ \Eprint {https://arxiv.org/abs/2206.08909} {arXiv:2206.08909
  [quant-ph]} \BibitemShut {NoStop}%
\bibitem [{\citenamefont {Zohar}\ and\ \citenamefont
  {Burrello}(2014)}]{Zohar2014}%
  \BibitemOpen
  \bibfield  {author} {\bibinfo {author} {\bibfnamefont {E.}~\bibnamefont
  {Zohar}}\ and\ \bibinfo {author} {\bibfnamefont {M.}~\bibnamefont
  {Burrello}},\ }\bibfield  {title} {\bibinfo {title} {A formulation of lattice
  gauge theories for quantum simulations},\ }\bibfield  {journal} {\bibinfo
  {journal} {Phys. Rev. D 91, 054506 (2015)}\ }\href
  {https://doi.org/10.1103/PhysRevD.91.054506} {10.1103/PhysRevD.91.054506}
  (\bibinfo {year} {2014}),\ \Eprint {https://arxiv.org/abs/1409.3085}
  {arXiv:1409.3085 [quant-ph]} \BibitemShut {NoStop}%
\bibitem [{\citenamefont {Zohar}\ and\ \citenamefont
  {Cirac}(2018)}]{Zohar2018}%
  \BibitemOpen
  \bibfield  {author} {\bibinfo {author} {\bibfnamefont {E.}~\bibnamefont
  {Zohar}}\ and\ \bibinfo {author} {\bibfnamefont {J.~I.}\ \bibnamefont
  {Cirac}},\ }\bibfield  {title} {\bibinfo {title} {Eliminating fermionic
  matter fields in lattice gauge theories},\ }\bibfield  {journal} {\bibinfo
  {journal} {Phys. Rev. B 98, 075119 (2018)}\ }\href
  {https://doi.org/10.1103/PhysRevB.98.075119} {10.1103/PhysRevB.98.075119}
  (\bibinfo {year} {2018}),\ \Eprint {https://arxiv.org/abs/1805.05347}
  {arXiv:1805.05347 [quant-ph]} \BibitemShut {NoStop}%
\bibitem [{\citenamefont {Mazzola}\ \emph {et~al.}(2021)\citenamefont
  {Mazzola}, \citenamefont {Mathis}, \citenamefont {Mazzola},\ and\
  \citenamefont {Tavernelli}}]{Mazzola2021}%
  \BibitemOpen
  \bibfield  {author} {\bibinfo {author} {\bibfnamefont {G.}~\bibnamefont
  {Mazzola}}, \bibinfo {author} {\bibfnamefont {S.~V.}\ \bibnamefont {Mathis}},
  \bibinfo {author} {\bibfnamefont {G.}~\bibnamefont {Mazzola}},\ and\ \bibinfo
  {author} {\bibfnamefont {I.}~\bibnamefont {Tavernelli}},\ }\bibfield  {title}
  {\bibinfo {title} {Gauge invariant quantum circuits for $u(1)$ and yang-mills
  lattice gauge theories},\ }\bibfield  {journal} {\bibinfo  {journal} {Phys.
  Rev. Research, 2021}\ }\href
  {https://doi.org/10.1103/PhysRevResearch.3.043209}
  {10.1103/PhysRevResearch.3.043209} (\bibinfo {year} {2021}),\ \Eprint
  {https://arxiv.org/abs/2105.05870} {arXiv:2105.05870 [quant-ph]} \BibitemShut
  {NoStop}%
\bibitem [{\citenamefont {Murairi}\ \emph {et~al.}(2022)\citenamefont
  {Murairi}, \citenamefont {Cervia}, \citenamefont {Kumar}, \citenamefont
  {Bedaque},\ and\ \citenamefont {Alexandru}}]{Murairi2022}%
  \BibitemOpen
  \bibfield  {author} {\bibinfo {author} {\bibfnamefont {E.~M.}\ \bibnamefont
  {Murairi}}, \bibinfo {author} {\bibfnamefont {M.~J.}\ \bibnamefont {Cervia}},
  \bibinfo {author} {\bibfnamefont {H.}~\bibnamefont {Kumar}}, \bibinfo
  {author} {\bibfnamefont {P.~F.}\ \bibnamefont {Bedaque}},\ and\ \bibinfo
  {author} {\bibfnamefont {A.}~\bibnamefont {Alexandru}},\ }\bibfield  {title}
  {\bibinfo {title} {How many quantum gates do gauge theories require?},\
  }\href@noop {} {\  (\bibinfo {year} {2022})},\ \Eprint
  {https://arxiv.org/abs/2208.11789} {arXiv:2208.11789 [hep-lat]} \BibitemShut
  {NoStop}%
\bibitem [{\citenamefont {Michael A.~Nielsen}(2010)}]{MichaelA.Nielsen2010}%
  \BibitemOpen
  \bibfield  {author} {\bibinfo {author} {\bibfnamefont {I.~L.~C.}\
  \bibnamefont {Michael A.~Nielsen}},\ }\href
  {https://doi.org/10.1017/cbo9780511976667} {\emph {\bibinfo {title} {Quantum
  Computation and Quantum Information}}}\ (\bibinfo {year} {2010})\BibitemShut
  {NoStop}%
\bibitem [{\citenamefont {{Qiskit contributors}}(2023)}]{Qiskit}%
  \BibitemOpen
  \bibfield  {author} {\bibinfo {author} {\bibnamefont {{Qiskit
  contributors}}},\ }\href {https://doi.org/10.5281/zenodo.2573505} {\bibinfo
  {title} {Qiskit: An open-source framework for quantum computing}} (\bibinfo
  {year} {2023})\BibitemShut {NoStop}%
\bibitem [{\citenamefont {Ortiz}\ \emph {et~al.}(2000)\citenamefont {Ortiz},
  \citenamefont {Gubernatis}, \citenamefont {Knill},\ and\ \citenamefont
  {Laflamme}}]{Ortiz2000}%
  \BibitemOpen
  \bibfield  {author} {\bibinfo {author} {\bibfnamefont {G.}~\bibnamefont
  {Ortiz}}, \bibinfo {author} {\bibfnamefont {J.~E.}\ \bibnamefont
  {Gubernatis}}, \bibinfo {author} {\bibfnamefont {E.}~\bibnamefont {Knill}},\
  and\ \bibinfo {author} {\bibfnamefont {R.}~\bibnamefont {Laflamme}},\
  }\bibfield  {title} {\bibinfo {title} {Quantum algorithms for fermionic
  simulations},\ }\href {https://doi.org/10.1103/physreva.64.022319} {\bibfield
   {journal} {\bibinfo  {journal} {Phys.Rev.A64:022319,2001}\ }\textbf
  {\bibinfo {volume} {64}},\ \bibinfo {pages} {022319} (\bibinfo {year}
  {2000})},\ \Eprint {https://arxiv.org/abs/cond-mat/0012334}
  {arXiv:cond-mat/0012334 [cond-mat]} \BibitemShut {NoStop}%
\bibitem [{\citenamefont {Somma}\ \emph {et~al.}(2002)\citenamefont {Somma},
  \citenamefont {Ortiz}, \citenamefont {Gubernatis}, \citenamefont {Knill},\
  and\ \citenamefont {Laflamme}}]{PhysRevA.65.042323}%
  \BibitemOpen
  \bibfield  {author} {\bibinfo {author} {\bibfnamefont {R.}~\bibnamefont
  {Somma}}, \bibinfo {author} {\bibfnamefont {G.}~\bibnamefont {Ortiz}},
  \bibinfo {author} {\bibfnamefont {J.~E.}\ \bibnamefont {Gubernatis}},
  \bibinfo {author} {\bibfnamefont {E.}~\bibnamefont {Knill}},\ and\ \bibinfo
  {author} {\bibfnamefont {R.}~\bibnamefont {Laflamme}},\ }\bibfield  {title}
  {\bibinfo {title} {Simulating physical phenomena by quantum networks},\
  }\href {https://doi.org/10.1103/PhysRevA.65.042323} {\bibfield  {journal}
  {\bibinfo  {journal} {Phys. Rev. A}\ }\textbf {\bibinfo {volume} {65}},\
  \bibinfo {pages} {042323} (\bibinfo {year} {2002})}\BibitemShut {NoStop}%
\bibitem [{\citenamefont {Zohar}(2019)}]{Zohar2019}%
  \BibitemOpen
  \bibfield  {author} {\bibinfo {author} {\bibfnamefont {E.}~\bibnamefont
  {Zohar}},\ }\bibfield  {title} {\bibinfo {title} {Local manipulation and
  measurement of nonlocal many-body operators in lattice gauge theory quantum
  simulators},\ }\href {https://doi.org/10.1103/physrevd.101.034518} {\bibfield
   {journal} {\bibinfo  {journal} {Phys. Rev. D 101, 034518 (2020)}\ }\textbf
  {\bibinfo {volume} {101}},\ \bibinfo {pages} {034518} (\bibinfo {year}
  {2019})},\ \Eprint {https://arxiv.org/abs/1911.11156} {arXiv:1911.11156
  [quant-ph]} \BibitemShut {NoStop}%
\bibitem [{\citenamefont {Borla}\ \emph {et~al.}(2019)\citenamefont {Borla},
  \citenamefont {Verresen}, \citenamefont {Grusdt},\ and\ \citenamefont
  {Moroz}}]{Borla2019}%
  \BibitemOpen
  \bibfield  {author} {\bibinfo {author} {\bibfnamefont {U.}~\bibnamefont
  {Borla}}, \bibinfo {author} {\bibfnamefont {R.}~\bibnamefont {Verresen}},
  \bibinfo {author} {\bibfnamefont {F.}~\bibnamefont {Grusdt}},\ and\ \bibinfo
  {author} {\bibfnamefont {S.}~\bibnamefont {Moroz}},\ }\bibfield  {title}
  {\bibinfo {title} {Confined phases of one-dimensional spinless fermions
  coupled to $z_2$ gauge theory},\ }\href
  {https://doi.org/10.1103/physrevlett.124.120503} {\bibfield  {journal}
  {\bibinfo  {journal} {Phys. Rev. Lett. 124, 120503 (2020)}\ }\textbf
  {\bibinfo {volume} {124}},\ \bibinfo {pages} {120503} (\bibinfo {year}
  {2019})},\ \Eprint {https://arxiv.org/abs/1909.07399} {arXiv:1909.07399
  [cond-mat.str-el]} \BibitemShut {NoStop}%
\bibitem [{\citenamefont {Mueller}\ \emph {et~al.}(2022)\citenamefont
  {Mueller}, \citenamefont {Zache},\ and\ \citenamefont {Ott}}]{Mueller2022}%
  \BibitemOpen
  \bibfield  {author} {\bibinfo {author} {\bibfnamefont {N.}~\bibnamefont
  {Mueller}}, \bibinfo {author} {\bibfnamefont {T.~V.}\ \bibnamefont {Zache}},\
  and\ \bibinfo {author} {\bibfnamefont {R.}~\bibnamefont {Ott}},\ }\bibfield
  {title} {\bibinfo {title} {Thermalization of gauge theories from their
  entanglement spectrum},\ }\href
  {https://doi.org/10.1103/physrevlett.129.011601} {\bibfield  {journal}
  {\bibinfo  {journal} {Physical Review Letters}\ }\textbf {\bibinfo {volume}
  {129}},\ \bibinfo {pages} {011601} (\bibinfo {year} {2022})}\BibitemShut
  {NoStop}%
\bibitem [{\citenamefont {Wang}\ \emph {et~al.}(2021)\citenamefont {Wang},
  \citenamefont {Li},\ and\ \citenamefont {Wang}}]{Wang2021}%
  \BibitemOpen
  \bibfield  {author} {\bibinfo {author} {\bibfnamefont {Y.}~\bibnamefont
  {Wang}}, \bibinfo {author} {\bibfnamefont {G.}~\bibnamefont {Li}},\ and\
  \bibinfo {author} {\bibfnamefont {X.}~\bibnamefont {Wang}},\ }\bibfield
  {title} {\bibinfo {title} {Variational quantum gibbs state preparation with a
  truncated taylor series},\ }\href
  {https://doi.org/10.1103/PhysRevApplied.16.054035} {\bibfield  {journal}
  {\bibinfo  {journal} {Phys. Rev. Appl.}\ }\textbf {\bibinfo {volume} {16}},\
  \bibinfo {pages} {054035} (\bibinfo {year} {2021})}\BibitemShut {NoStop}%
\end{thebibliography}%

\onecolumngrid

\appendix

\end{document}